\documentclass[a4paper,11pt]{article}
\usepackage{jinstpub} 
\usepackage{lineno}
\usepackage[nolist]{acronym} 
\usepackage{gensymb}
\usepackage{comment}
\usepackage{subcaption}
\usepackage{siunitx} 
\usepackage{orcidlink} 

\title{\boldmath Study of the intra-pixel timing resolution of planar silicon sensors using the TDCpix ASIC}

\author[g,1]{A.~Shepelev~\orcidlink{0000-0002-8993-5069},\note{Corresponding authors.}}
\author[g,1]{C.~Sam~\orcidlink{0009-0005-3802-5777},}

\author[a]{B.~Velghe~\orcidlink{0000-0002-0797-8381},}

\author[b]{M.~Perrin-Terrin~\orcidlink{0000-0002-3568-1956},}

\author[c]{K.~Akiba~\orcidlink{0000-0002-6736-471X},} 
\author[c]{M.~van~Beuzekom~\orcidlink{0000-0002-0500-1286},} 
\author[c]{T.~Evans~\orcidlink{0000-0003-3016-1879},} 
\author[c]{K.~Heijhoff~\orcidlink{0000-0001-5407-7466},} 

\author[d]{Z.~Kucerova~\orcidlink{0000-0001-8906-3902},}

\author[e]{V.~Coco~\orcidlink{0000-0002-5310-6808},}  
\author[e]{R.~Dumps,} 
\author[e]{E.~Lemos~Cid~\orcidlink{0000-0003-3001-6268},} 
\author[e]{M.~Noy~\orcidlink{0009-0001-3394-3019},}
\author[e]{T.~Pajero~\orcidlink{0000-0001-9630-2000},} 
\author[e]{E.~Rodríguez~Rodríguez~\orcidlink{0000-0002-7973-8061},} 
\author[e]{M.~Williams~\orcidlink{0000-0001-7765-8941},} 

\author[f]{A.~Kleimenova~\orcidlink{0000-0002-9129-4985},}

\author[g]{K.~Krizka~\orcidlink{0000-0001-6408-2648},}
\author[g]{C.~Lazzeroni~\orcidlink{0000-0003-4074-4787},}
\author[g,2]{M.~Madurai~\orcidlink{0000-0002-6503-0759},\note{Present address: University of Warwick, Warwick, United Kingdom}} 

\author[h]{D.~Bacher~\orcidlink{0000-0002-1249-367X}} 

\affiliation[a]{TRIUMF, \\Vancouver, British Columbia, Canada}
\affiliation[b]{Aix Marseille University, CNRS/IN2P3, CPPM,\\Marseille, France}
\affiliation[c]{Nikhef, \\Amsterdam, the Netherlands}
\affiliation[d]{Comenius University,\\Bratislava, Slovakia}
\affiliation[e]{CERN, European Organization for Nuclear Research,\\Geneva, Switzerland}
\affiliation[f]{École Polytechnique Fédérale de Lausanne,\\Lausanne, Switzerland}
\affiliation[g]{University of Birmingham,\\Birmingham, United Kingdom}
\affiliation[h]{University of Oxford, \\Oxford, United Kingdom}

\emailAdd{a.shepelev@bham.ac.uk}
\emailAdd{ceyhan.sam@cern.ch}

\abstract{A study of the intra-pixel timing resolution in planar silicon detectors, bonded to the TDCpix \ac{ASIC}, is performed using the TimePix4 telescope. Spatial non-uniformities in the timing resolution, arising from the weighting field, are measured for the first time in an experimental environment. The results are validated using simulation and discussed in the context of future \ac{RD} goals.}

\keywords{particle tracking detectors (solid-state detectors), hybrid detectors, timing detectors}

\begin{acronym}
    \acro{GTK}{GigaTracKer}
\end{acronym}

\begin{acronym}
    \acro{HIKE}{High Intensity Kaon Experiments}
\end{acronym}

\begin{acronym}
    \acro{HL-LHC}{High Luminosity Large Hadron Collider}
\end{acronym}

\begin{acronym}
    \acro{SPS}{Super Proton Synchrotron}
\end{acronym}

\begin{acronym}
    \acro{NA}{North Area}
\end{acronym}

\begin{acronym}
    \acro{LS3}{Long Shutdown 3}
\end{acronym}

\begin{acronym}
    \acro{LS4}{Long Shutdown 4}
\end{acronym}

\begin{acronym}
    \acro{VELO}{VErtex LOcator}
\end{acronym}

\begin{acronym}
    \acro{ASIC}{Application Specific Integrated Circuit}
\end{acronym}

\begin{acronym}
    \acro{ToT}{Time-over-Threshold}
\end{acronym}

\begin{acronym}
    \acro{ToA}{Time of Arrival}
\end{acronym}

\begin{acronym}
    \acro{TDC}{Time-to-Digital Converter}
\end{acronym}

\begin{acronym}
    \acro{DAC}{Digital-to-Analogue Converter}
\end{acronym}

\begin{acronym}
    \acro{DLL}{Delay-Locked Loop}
\end{acronym}

\begin{acronym}
    \acro{EoC}{End of Column}
\end{acronym}

\begin{acronym}
    \acro{QC}{Quarter-Chip}
\end{acronym}

\begin{acronym}
    \acro{TP}{Test Pulse}
\end{acronym}

\begin{acronym}
    \acro{SLVS}{Scalable Low Voltage Signaling}
\end{acronym}

\begin{acronym}
    \acro{LVDS}{Low Voltage Differential Signaling}
\end{acronym}

\begin{acronym}
    \acro{SNR}{Signal-to-Noise}
\end{acronym}

\begin{acronym}
    \acro{MPV}{Most Probable Value}
\end{acronym}

\begin{acronym}
    \acro{DUT}{Device Under Test}
\end{acronym}

\begin{acronym}
    \acro{PCB}{Printed Circuit Board}
\end{acronym}

\begin{acronym}
    \acro{MAPS}{Monolithic Active Pixel Sensors}
\end{acronym}

\begin{acronym}
    \acro{LGAD}{Low Gain Avalanche Diode}
\end{acronym}

\begin{acronym}
    \acro{CSA}{Charge Sensistive Amplifier}
\end{acronym}

\begin{acronym}
    \acro{MCP}{Micro-Channel Plate Photo-Multiplier}
\end{acronym}

\begin{acronym}
    \acro{ToF}{Time-of-Flight}
\end{acronym}

\begin{acronym}
    \acro{MIP}{Minimum Ionizing Particle}
\end{acronym}

\begin{acronym}
    \acro{PAI}{Photo Absorption Ionization}
\end{acronym}

\begin{acronym}
    \acro{HEP}{High Energy Physics}
\end{acronym}

\begin{acronym}
    \acro{RD}[R\&D]{Research and Development}
\end{acronym}

\begin{acronym}
    \acro{TLU}{Telescope Logic Unit}
\end{acronym}

\begin{acronym}
    \acro{FPGA}{Field Programmable Gate Array}
\end{acronym}

\begin{acronym}
    \acro{MPV}{Most Probable Value}
\end{acronym}
\begin{document}
\maketitle
\flushbottom

\section{Introduction}
\label{sec:Introduction}
The timing resolution of 4D tracking systems is paramount for current generation experiments to achieve their targeted physics goals. The NA62 experiment, which aims to measure the $K^+ \rightarrow \pi^+\nu\bar{\nu}$ branching fraction to 15\% precision, requires a timing resolution on the $K^+$ track of $\mathcal{O}$(\qty{100}{\ps}) \cite{NA62DetectorPaper}. The NA62 4D beam tracker \ac{GTK}, composed of four hybrid silicon pixel detectors, provides a single hit timing resolution of \qty{130}{\ps} and a track timing resolution of \qty{65}{\ps} \cite{GTKPrototypeResultsPaper, GigaTrackerPaper_2017, NA62GTKPaper, GigaTrackerPaper_2020}.
\paragraph{}
Furthermore, the development of 4D tracking systems with fast timing performance has been identified as a key goal in the \ac{HEP} community \cite{ECFARoadmap}. For example, to cope with the high rate environment foreseen in the \ac{HL-LHC} after \ac{LS4}, the LHCb Upgrade II targets a single hit timing resolution of \qty{<= 50}{\ps} in the \ac{VELO} 2 system \cite{LHCbUpragde2Paper}. Similar timing requirements have been reported for the timing layer upgrades in ATLAS \cite{ATLASUpgradeReport} and CMS \cite{CMSUpgradeReport}, and the nuSCOPE beam spectrometer \cite{nuSCOPEPaper}.
\paragraph{}
In this report, the timing resolution of planar silicon sensors bonded to the TDCpix \ac{ASIC} \cite{TDCpixPaper, TDCpixProceedings, TDCPixCharacterisation, KLUGE2013511}, the \ac{DUT}, is studied using the TimePix4 telescope \cite{akiba2025timepix4beamtelescope}. TDCpix was developed for the read-out of silicon pixel detectors, originally for use in the NA62 \ac{GTK}. A Python framework for analysing TDCpix data and associating TimePix4 tracks was created using the NumPy \cite{NumPyPaper} and SciPy \cite{2020SciPy-NMeth} packages. Non-uniformities in the intra-pixel timing resolution are studied using the \ac{ToA} of beam particles and discussed in the context of the pixel electrode weighting field. The results are validated using a data-driven model and an AllPix$^2$ simulation \cite{AllPix2}. Finally, the impact of this work on current and future \ac{RD} goals is discussed. 

\section{Device Under Test}
\label{sec:Device-Under-Test}
\subsection{TDCpix}
\label{subsec:ASIC-Overview}
The TDCpix \ac{ASIC} is a data-driven (trigger-less) device designed to precisely measure the \ac{ToA} of charged particles (hits) using in-pixel \ac{ToT} discriminators and on-chip \acp{TDC}. For each hit, the \ac{ASIC} records the pixel address, the leading edge and the trailing edge times of the signal (from which the \ac{ToT} is calculated). TDCpix consists of 1800 pixels, each measuring \qtyproduct[product-units=single]{300 x 300}{\um\squared}, arranged in a matrix of 40 columns and 45 rows. The 40 columns are grouped into 20 column pairs, with each pair sharing a \ac{DLL} for timing reference. To reduce the number of required \acp{TDC}, five non-adjacent pixels in a column share a single \ac{TDC}.
The \ac{TDC} provides a 32-bit timestamp with a fine time binning of $\simeq$ \qty{100}{\ps}\footnote{In practice, the time binning is \qty{97.62625}{\ps}, when operated at the nominal \ac{DLL} frequency.}, when operated at the nominal \ac{DLL} frequency of \qty{320}{\MHz}. Data readout is performed by one of four independent readout units, the \ac{QC}, each serving a set of 10 columns. The TDCpix specifications are summarised in Table \ref{tab:spec}, while a complete description of the \ac{ASIC} design and architecture can be found in \cite{TDCpixPaper}.

\begin{table}[htbp]
\centering
\caption{TDCpix specifications\label{tab:spec}}
\smallskip
\begin{tabular}{l|c}
\hline
Specification&Value \\
\hline
Number of pixels per chip & 1800 = \numproduct{45 x 40} \\
Pixel size & \qtyproduct[product-units=single]{300 x 300}{\um\squared} \\
Active area per chip & \qtyproduct{12 x 13.5}{\mm} = \qty{162}{\mm\squared} \\
ASIC time binning & $\simeq$ \qty{100}{\ps} \\
Thickness of read-out chip & \qty{100}{\um} \\
Dynamic input range & \qtyrange[range-phrase=--, range-units=single]{0.7}{10}{\femto\coulomb}, \qtyrange[range-phrase=--, range-units=single]{5000}{60000}{\elementarycharge} \\
Design particle rate per chip & \qty{105}{\MHz} \\
Average rate per pixel & \qty{58}{\kHz} \\
\hline
\end{tabular}
\end{table}
\subsection{Sensor}
\label{subsec:Sensor}
A series of \qty{200}{\um} thick n-on-p planar silicon sensors, provided by the CERN EP \ac{RD} programme WP 1.1 on hybrid silicon pixel detectors, were bump-bonded to the TDCpix \ac{ASIC}, collectively referred to as hybrids. The sensors were developed and manufactured by ADVACAM using non-oxygenated wafers with a bulk resistivity of approximately \qty{10}{\kohm\cm}. A detailed study and characterisation of the sensors can be found in \cite{JakobHaimbergerThesis}. The hybrids were wire-bonded to a custom \ac{PCB}, the daughter-card. The sensors could be biased up to $\approx$ \qty{400}{\V} before showing signs of breakdown. Therefore, during testing, the maximum applied bias voltage was \qty{300}{\V}.

\section{Testbeam}
\label{sec:Testbeam}
\subsection{TimePix4 Telescope}
\label{subsec:TimePix4-Telescope}
\begin{figure}[h]
    \centering
    \begin{subfigure}[t]{0.45\textwidth}
        \centering
        \includegraphics[width=0.7\textwidth, angle=0]{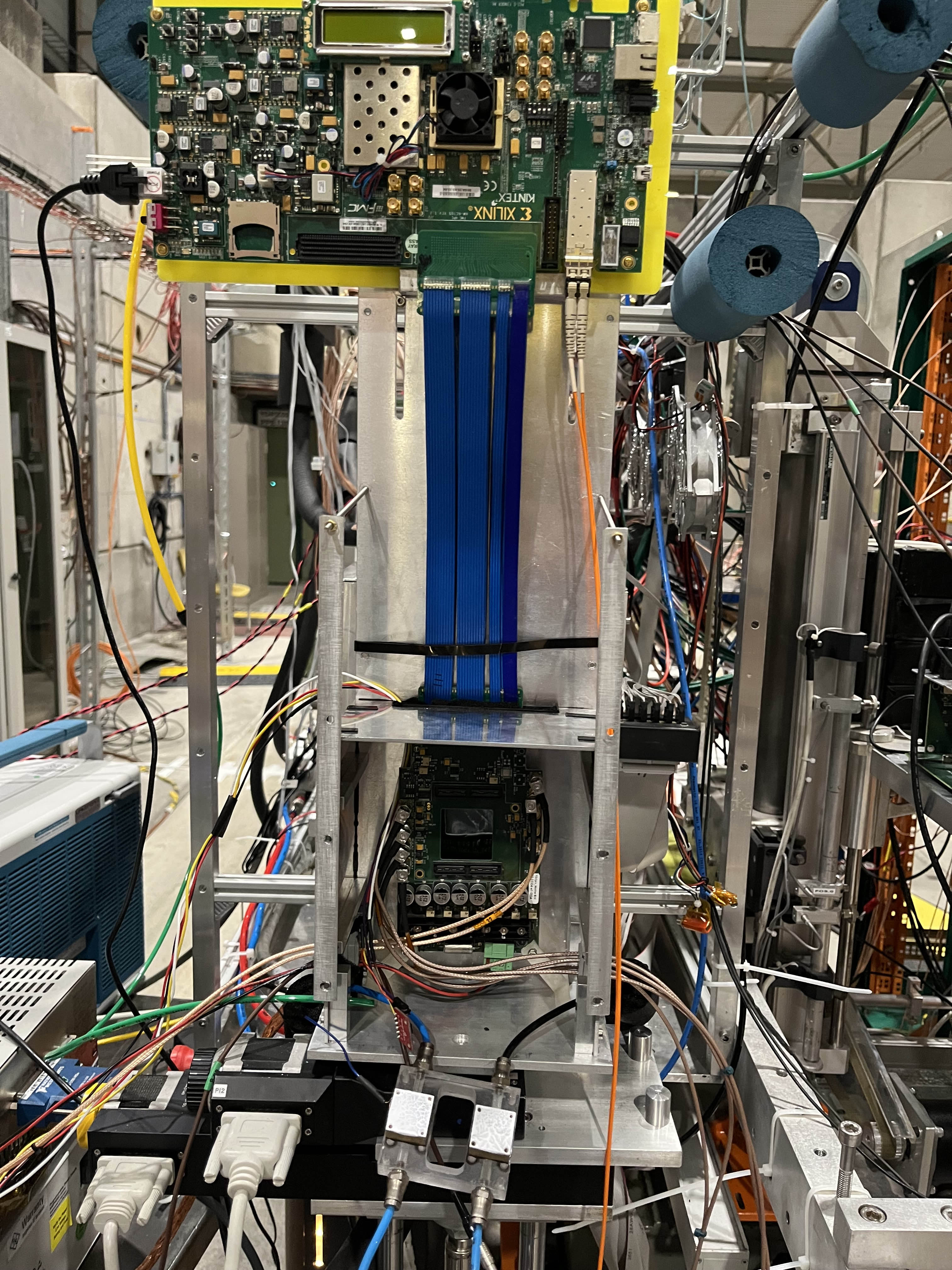}
        \caption{}
        \label{fig:TDCPix_DUT_Open}
    \end{subfigure}
    \begin{subfigure}[t]{0.45\textwidth}
        \centering
        \includegraphics[width=0.7\textwidth, angle=0]{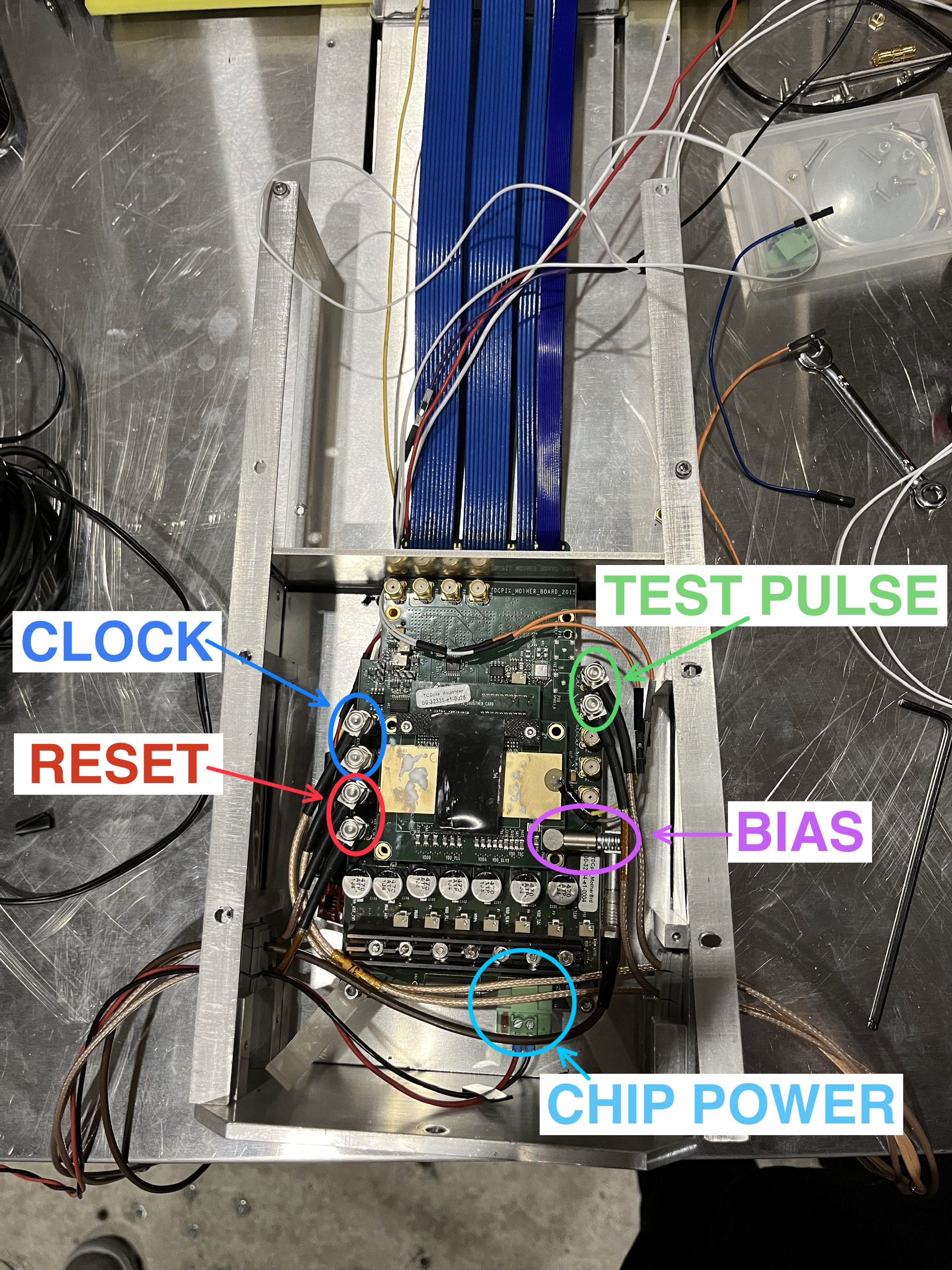}
        \caption{}
        \label{fig:TDCPix_DUT_Closed}
    \end{subfigure}
    \caption{(a) A photo of the open \ac{DUT} box installed into the TimePix4 telescope, taken from the downstream end of the telescope. The AMD Xilinx Kintex KC705 \ac{FPGA} can be seen above the \ac{DUT}. (b) Close up of a single TDCpix hybrid inside the \ac{DUT} box, the active area of the sensor is protected by a black cover. The cables providing the \ac{DLL} clock (blue), reset (red), test-pulse signals (green), power (cyan) and sensor bias voltage (pink) can be seen.}
    \label{fig:TDCPix-Integration}
\end{figure}

The TDCpix hybrids were studied using the TimePix4 telescope, located on the H8 beamline in the North Area of the CERN \ac{SPS}. The H8 beamline provides a \qty{180}{\GeV} mixed hadron beam, derived from the 400~GeV primary \ac{SPS} beam, in 4.8~s extractions (spills). The TimePix4 telescope consists of eight tracking planes that offer, at the central stage, a track time resolution of \qty{92.5\pm5}{\ps} and a spatial resolution of \qty{2.3\pm0.1}{\um}  (\qty{2.4\pm0.1}{\um}) in the $x$ ($y$) direction \cite{akiba2025timepix4beamtelescope}. The \ac{DUT} was mounted on a three-axis rotation stage, \qty{900}{\mm} downstream of the centre of the telescope, inside an environmentally isolated \ac{DUT} box, shown in Figure \ref{fig:TDCPix-Integration}. At this position the track spatial resolution degrades to \qty{12.0\pm0.5}{\um}. The spatial resolution of the TimePix4 telescope is determined via simulation as a function of $z$, further details on the procedure can be found in \cite{akiba2025timepix4beamtelescope}.

\subsection{Telescope Integration Scheme}
\label{subsec:Telescope-Integration-Scheme}
\begin{figure}[h]
    \centering
    \includegraphics[width=\linewidth]{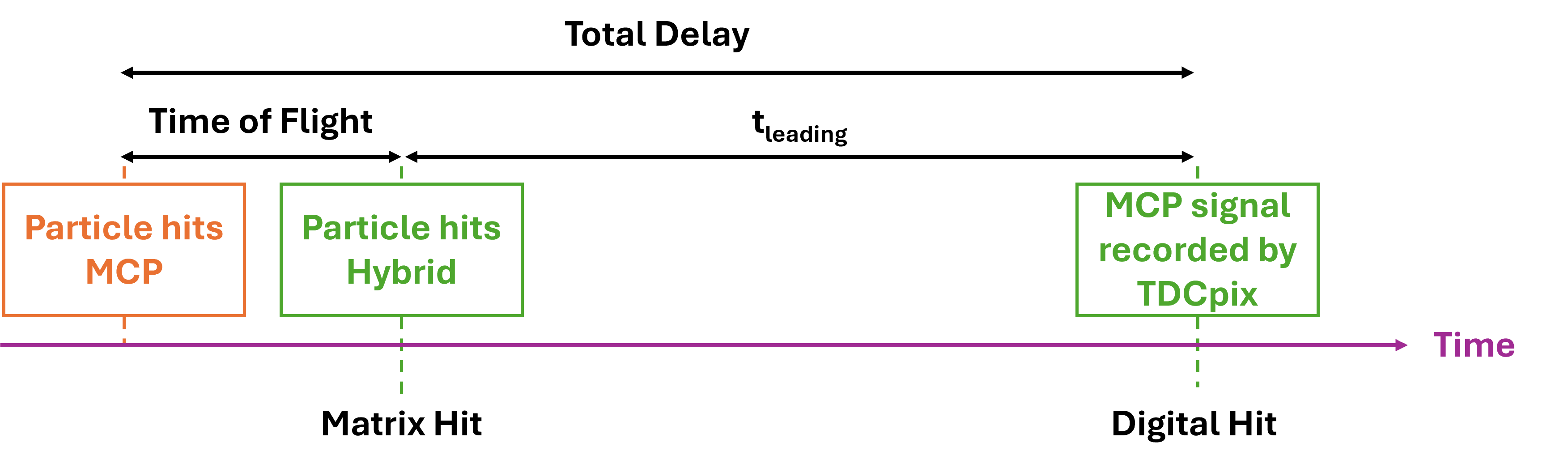}
    \caption{Qualitative timing diagram illustrating the relative time of the \ac{MCP} signal, TDCpix matrix hit and TDCpix digital hit. Relative distances are not representative of accurate time differences and are purely illustrative.}
    \label{fig:TimePix4-Timing-Integration}
\end{figure}

\begin{figure}[h]
    \centering
    \includegraphics[width=\linewidth]{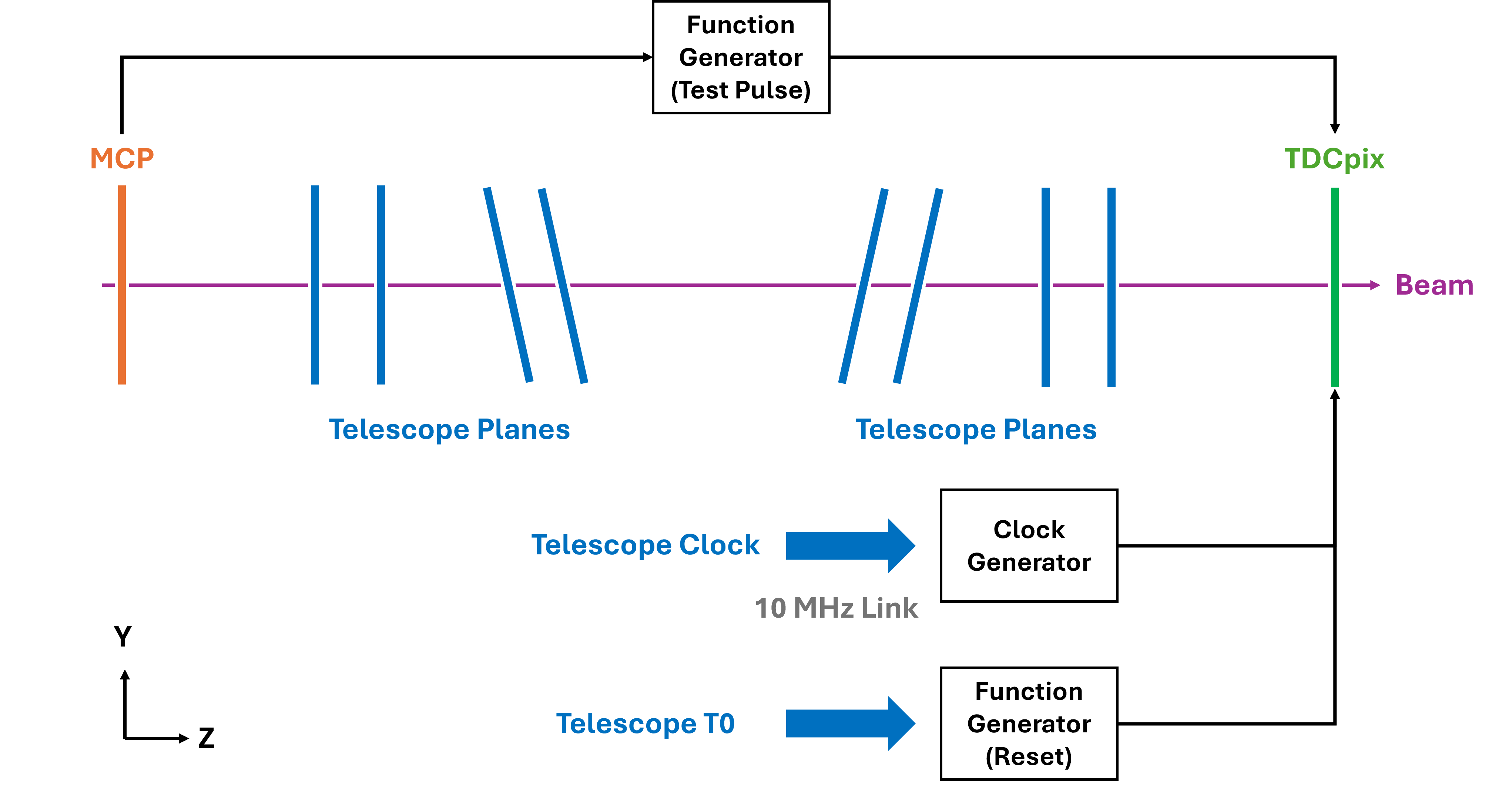}
    \caption{Simplified block diagram of TDCpix integration into the TimePix4 telescope environment (not to scale). See Section \ref{subsec:TDCPix-DAQ-Readout} for the definition of the Telescope T0.}
    \label{fig:TimePix4-Integration}
\end{figure}

A Photonis PP2365Y \ac{MCP}, already integrated into the TimePix4 environment \cite{akiba2025timepix4beamtelescope} and located upstream of the telescope, acted as a timing reference for TDCpix. The \ac{MCP} signals were delivered and timestamped by TDCpix through an intermediate Active Technologies PG1072 pulse generator, as TDCpix requires a differential input. The single-ended \ac{MCP} output signal was delivered, via an intermediate PicoTDC board \cite{PicoTDC}, to the trigger input of the PG1072 which subsequently generated a suitable differential DC-coupled \ac{LVDS} output. The final signal is routed to an arbitrary TDCpix \ac{TDC}, located at the periphery of the matrix. This sequence is hereinafter referred to as the time-stamping chain. This configuration leads to two signals recorded in the TDCpix output for every particle that traverses the telescope environment: one from the particle interacting with the sensitive detector volume, the ``matrix hit'' and one from the external \ac{MCP} signal, the ``digital hit''. The time difference between these two signals includes: the time of flight between the \ac{MCP} and \ac{DUT}, the signal propagation delay between the \ac{MCP} output and TDCpix input and other constant delays introduced by the PicoTDC board and PG1072. A qualitative diagram illustrating the timing scheme between the \ac{MCP} and TDCpix is shown in Figure \ref{fig:TimePix4-Timing-Integration}. The total contribution to the timing resolution from the \ac{MCP} and PicoTDC board is quoted to be \qty{12}{\ps} \cite{akiba2025timepix4beamtelescope}. Similarly, the expected contribution from the time-stamping chain was \qty{<41}{\ps}; consisting of the PG1072 trigger-in-to-output RMS jitter, listed as \qty{<30}{\ps} \cite{PG1072-Pulse-Generator}, and the \ac{TDC} quantisation error of \qty{97.7}{\ps} $/\sqrt{12} = $ \qty{28}{\ps}. 
\paragraph{}
The clock for TDCpix was provided by an external clock generator, synchronised to the TimePix4 telescope through a \qty{10}{\MHz} reference link. The telescope integration scheme is illustrated in Figure \ref{fig:TimePix4-Integration}. The temperature of the hybrid was monitored using a PT100 temperature sensor, thermally coupled to the hybrid, and controlled using a Peltier cooling system. The temperature of the hybrid was maintained around \qty{20}{\degreeCelsius} using this setup. Similarly, the humidity was monitored using a HIH-6130 sensor and maintained with a constant flow of dry air ($\mathcal{O}(1\%)$ relative humidity) into the \ac{DUT} box.

\subsection{Data Acquisition and Readout}
\label{subsec:TDCPix-DAQ-Readout}
The TDCpix data acquisition is performed in a continuous manner and the readout is entirely data driven. An AMD Xilinx Kintex KC705 \ac{FPGA} provided a control interface for TDCpix and the data readout. The TDCpix data stream consists of two primary outputs: data words, recording the hit and timing information; and frame words, used to increase the dynamic range of the \acp{TDC}. When a hit above threshold is registered, the leading time, \ac{ToT} and pixel address are encoded into a 48 bit data word and are written to the output stream. In tandem, every 2048 \ac{DLL} clock cycles (\qty{6.4}{\us} at the nominal clock frequency) a 28 bit frame counter is increased and a frame word is written, recording the frame counter and number of hits recorded in the frame. The frame counter continually increments until it rolls over every \qty{\sim29}{\minute} or until an external reset signal is received, after which it increments from zero. The data output stream is sent from TDCpix to the \ac{FPGA} and finally to the storage PC, where it is continuously written to a byte-file. The data were collected in extended periods denoted as runs; where a run consists of \numrange[range-phrase=--]{10}{100} spills from the \ac{SPS} and corresponds to a specific bias voltage applied to the sensor. Each run was initiated by an external \ac{LVDS} reset signal, the ``Telescope T0'', generated by the TimePix4 \ac{TLU} \cite{akiba2025timepix4beamtelescope}.
\paragraph{}
The aforementioned time synchronization scheme produced pairs of data words in the output data stream for every particle interaction, corresponding to the matrix and digital (\ac{MCP}) hits. As the digital hit corresponds to a \ac{TDC} on the periphery of the pixel matrix, and hence spatially separated from the beam, the pairs could be readily identified in the TDCpix output. The time difference between paired matrix and digital hits was approximately \qty{170}{\ns}, which is a constant offset that does not impact the analysis procedure. The TDCpix and TimePix4 data streams, consisting of the raw spatial and timing information, were written independently and required offline event matching. The TimePix4 data was reconstructed using the Kepler framework, details can be found in \cite{akiba2025timepix4beamtelescope}, and the resulting tracks were matched to the TDCpix data using the procedure described in Section \ref{sec:TimePix4-Track-Integration}.

\section{Timing Resolution}
\label{sec:Hybrid-Timing-Performance}
Initially, the timing resolution of the hybrid was studied and compared against data collected with the NA62 \ac{GTK}. Then, the \ac{ASIC} and external contributions were subtracted to determine the timing resolution of the sensor, which is compared against a theoretical model presented in \cite{TimeResolutionOfSiliconPixelSensors}.

\subsection{Charge Deposition}
\label{subsec:Charge-Deposition}
\begin{figure}
    \centering
    \begin{subfigure}[t]{0.49\textwidth}
        \centering
        \includegraphics[width=\textwidth]{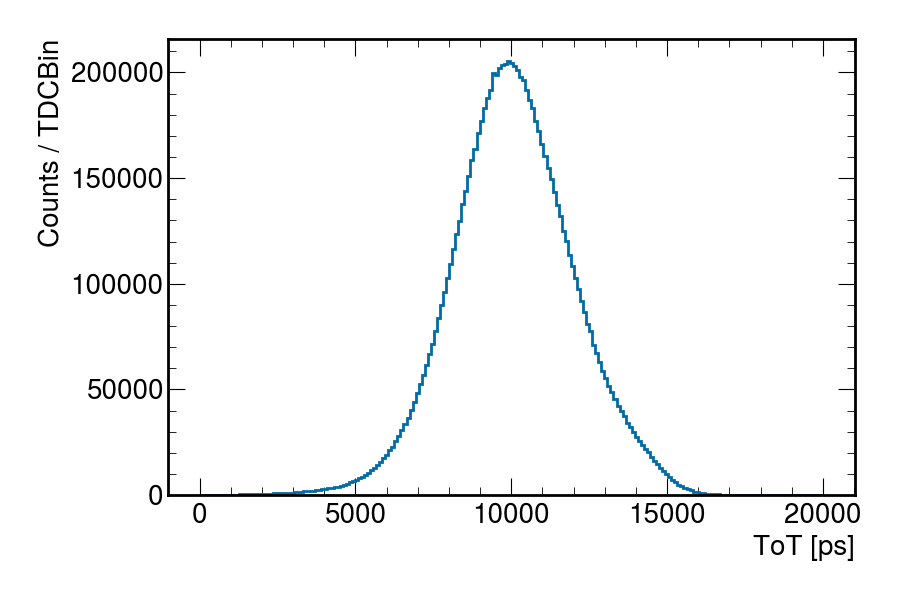}
        \caption{}
        \label{fig:ToT-Distribution-200V}
    \end{subfigure}
    \hfill
    \begin{subfigure}[t]{0.47\textwidth}
        \includegraphics[width=\textwidth]{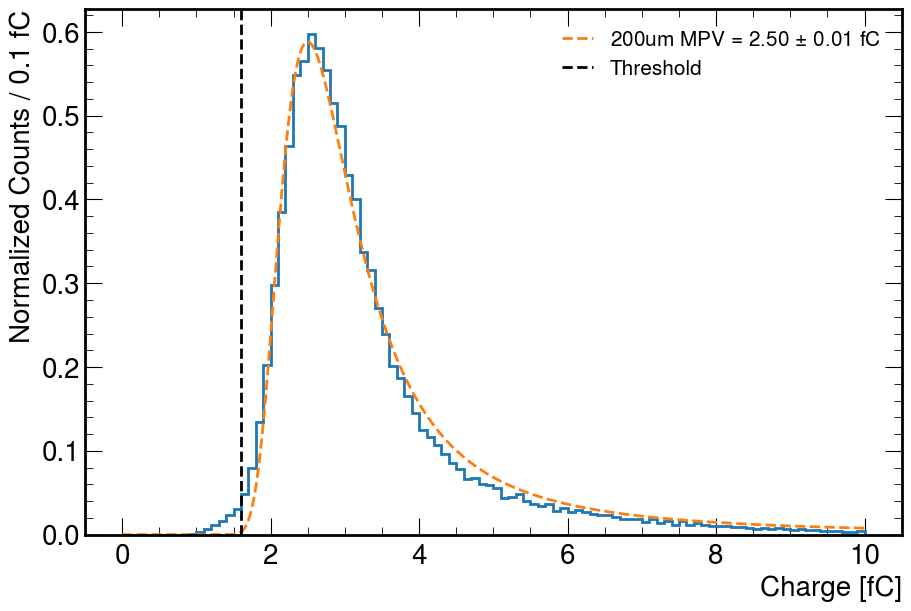}
        \caption{}
        \label{fig:Reconstructed-Charge-Distribution-200V}
    \end{subfigure}
    \caption{(a) \acl{ToT} distribution obtained at a bias voltage of \qty{200}{\V}. (b) The associated charge distribution, reconstructed using fixed charge injections. A fitted Landau distribution is overlaid (orange) and the \ac{MPV}, extracted from the fit, is stated. The associated error is the uncertainty on the fit parameter. The \qty{1.6}{\femto\coulomb} threshold is indicated by the dashed black line. }
    \label{fig:placeholder}
\end{figure}

TDCpix operates with a fixed, pixel-dependent, discriminator threshold which needed to be optimised for charge depositions from \acp{MIP}. The expected \ac{MPV} of charge deposition for a \ac{MIP} in \qty{200}{\um} of silicon is \qty{2.4}{\femto\coulomb} \cite{PDG_2006}, therefore a threshold of \qty{1.6}{\femto\coulomb} was chosen. An example \ac{ToT} distribution (a proxy for deposited charge) obtained at a bias voltage of \qty{200}{\V} is shown in Figure \ref{fig:ToT-Distribution-200V}. The distribution does not reflect the expected Landau distribution due to the non-linear response of the pixel front-end circuit, consisting of a \ac{CSA} and shaper circuit \cite{TDCpixPaper}. The underlying charge distribution was obtained by performing a \ac{ToT} calibration. A series of fixed charge injections were performed at various charges and the average \ac{ToT} was extracted from a Gaussian fit. The resulting charge distribution is shown in Figure \ref{fig:Reconstructed-Charge-Distribution-200V}, displaying the expected Landau distribution. The obtained \ac{MPV} was \qty{2.5}{\femto\coulomb}, compatible with the expected charge deposition.

\subsection{Measurement of Timing Resolution}
\label{subsec:Approach-to-Time-Resolution-Characterization}
\begin{figure}[h]
    \centering
    \begin{subfigure}[t]{0.49\textwidth}
        \centering
        \includegraphics[width=\textwidth]{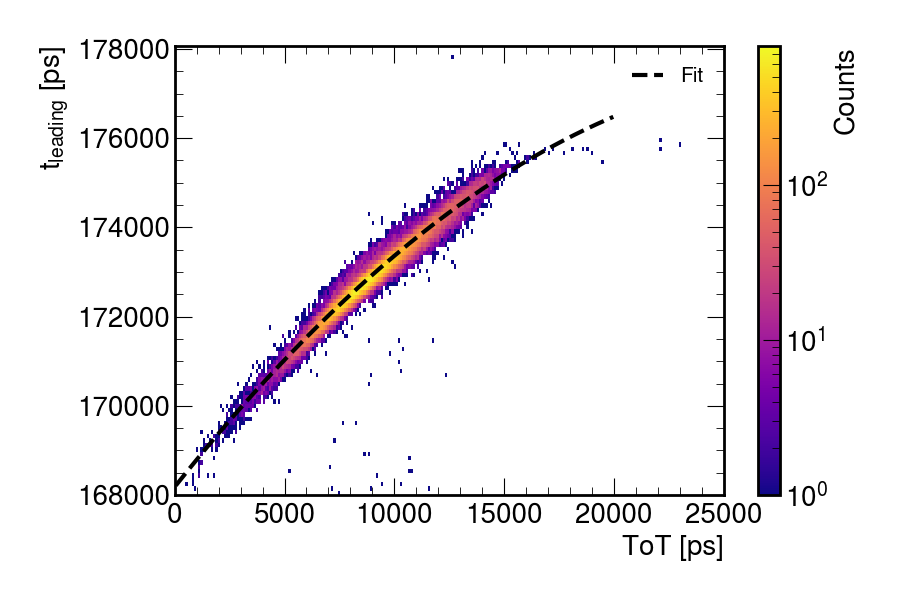}
        \caption{}
        \label{fig:Pixel-Timewalk-Curve}
    \end{subfigure}
    \hfill
    \begin{subfigure}[t]{0.49\textwidth}
        \centering
        \includegraphics[width=\textwidth]{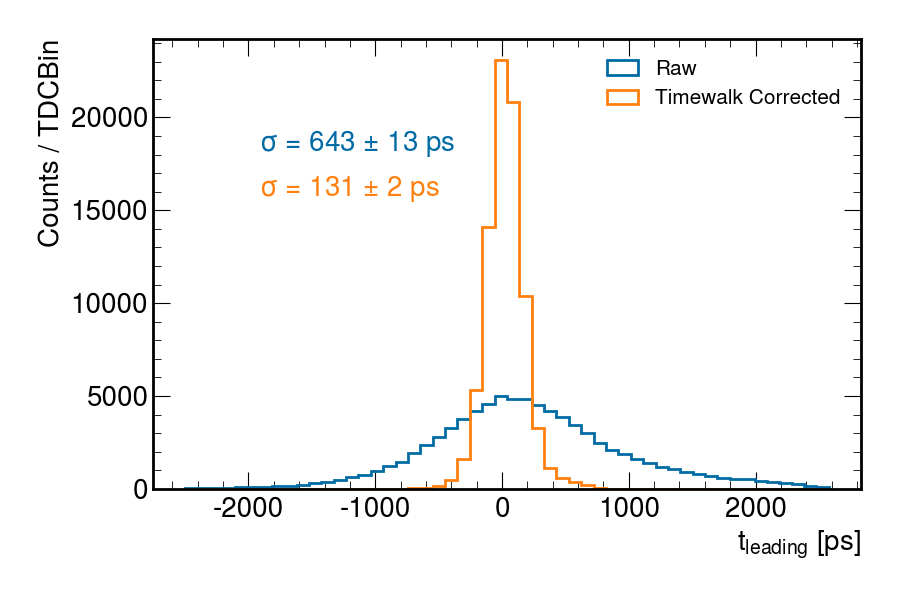}
        \caption{}
        \label{fig:Pixel-Timing-Distribution}
    \end{subfigure}
    \caption{(a) $t_\text{leading}$ vs \ac{ToT} for a single pixel fit with equation \ref{eq:Time-Walk-Fit}, illustrating the time-walk effect. (b) The timing distribution of the same pixel before (blue) and after (orange) the time-walk correction was applied. The standard deviation of each distribution, evaluated from a Gaussian fit, is stated. The mean of the raw timing distribution (blue) has been shifted to zero for comparison.}
\end{figure}

\begin{figure}
    \begin{subfigure}[t]{0.49\textwidth}
        \centering
        \includegraphics[width=\textwidth]{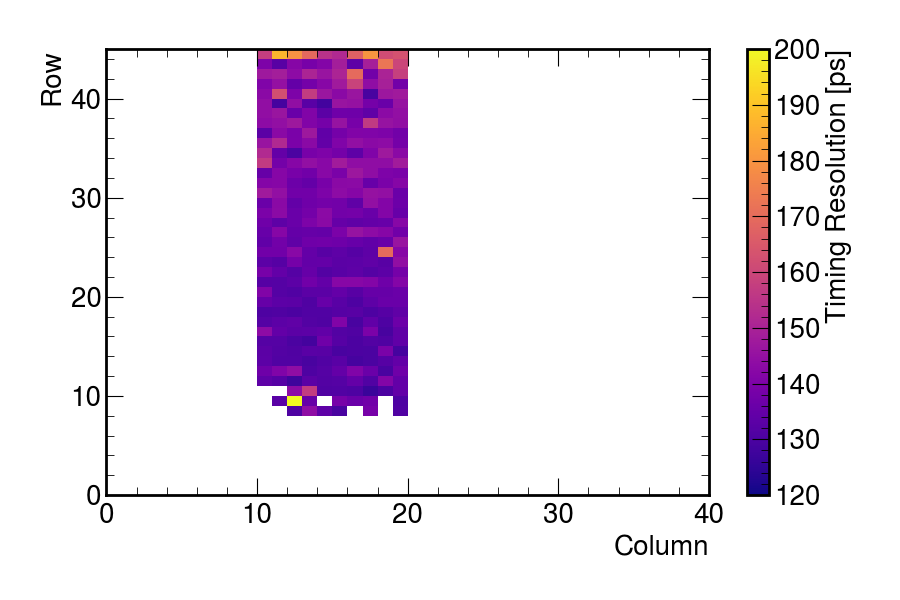}
        \caption{}
        \label{fig:Timing-Resolution-Map-200V}
    \end{subfigure}
    \hfill
    \begin{subfigure}[t]{0.49\textwidth}
        \centering
        \includegraphics[width=\textwidth]{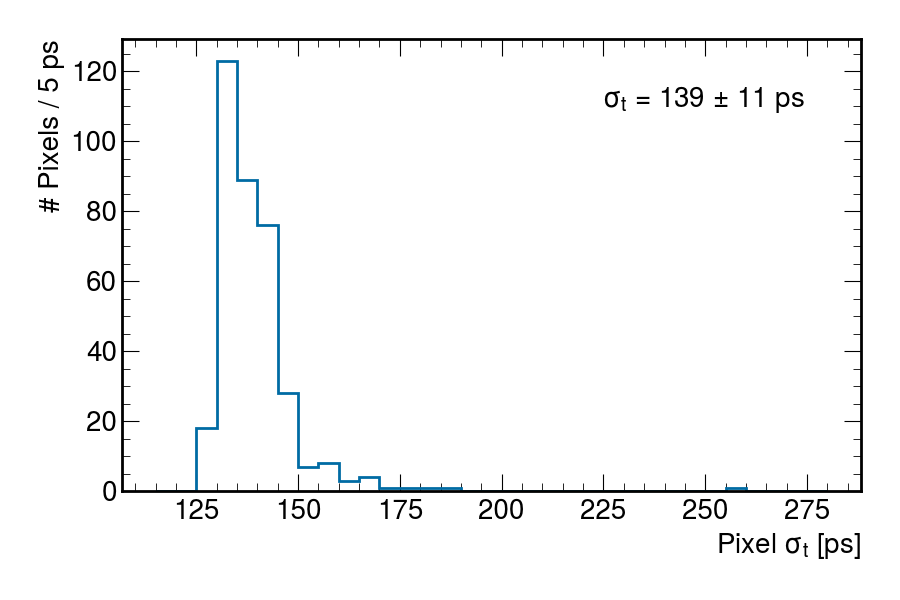}
        \caption{}
        \label{fig:Matrix-Pixel-Time-Resolution}
    \end{subfigure}
    \caption{(a) 2D map of the time-walk corrected pixel-by-pixel timing resolution obtained at a bias voltage of \qty{200}{\V}. Pixels with less than \num{500} entries were excluded from the timing resolution calculation and the color scale has been fixed to [120, 200] \unit{\ps}. (b) Associated 1D distribution of the pixel-by-pixel timing resolution map. The mean of this histogram gives the pixel average timing resolution, which is stated alongside the standard deviation. Only \ac{QC} 1 was enabled during this run.}
    \label{fig:Pixel-Average-Timing-Resolution-200V}
\end{figure}

\begin{figure}
    \begin{subfigure}[t]{0.49\textwidth}
        \centering
        \includegraphics[width=\textwidth]{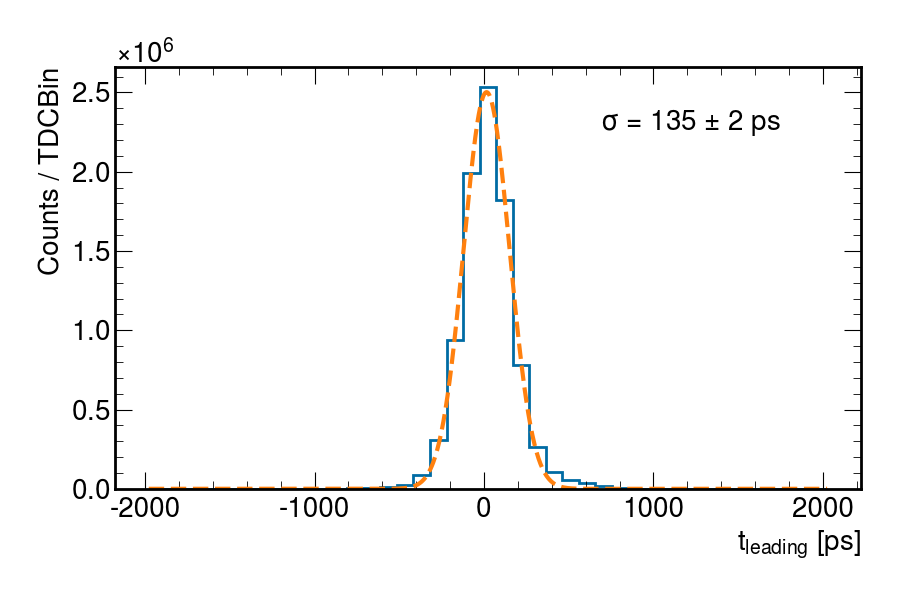}
        \caption{}
        \label{fig:Matrix-Time-Resolution}
    \end{subfigure}
    \hfill
    \begin{subfigure}[t]{0.47\textwidth}
        \centering
        \includegraphics[width=\textwidth]{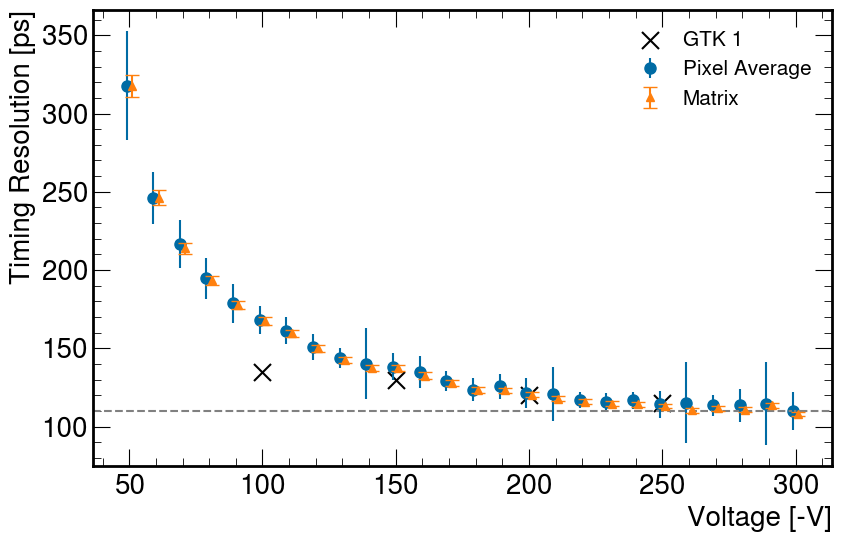}
        \caption{}
        \label{fig:Hybrid-Time-Resolution}
    \end{subfigure}
    \caption{(a) Example of the matrix timing resolution distribution, obtained for a sensor bias voltage of \qty{200}{\V}. The matrix timing resolution is extracted from a Gaussian fit (dashed orange line), the error represents the uncertainty on the fit parameter. (b) The hybrid timing resolution as a function of the applied bias voltage (obtained at T = \qty{20\pm0.2}{\degreeCelsius}), compared with data taken from NA62 \ac{GTK}1 (obtained at T \qty{<5}{\degreeCelsius}) \cite{NA62GTKPaper}. The pixel average (blue circles) and matrix (orange triangles) timing resolution values are plotted and their markers offset by \qty{-1}{\V} and \qty{+1}{\V}, respectively, for readability.}
    \label{fig:placeholder}
\end{figure}

Variations in the charge deposition result in a time-walk, degrading the timing resolution of the hybrid. Therefore, an offline correction was performed to recover the timing performance. The dependence of the leading-edge time on the \ac{ToT} was approximated by a quadratic function of the form: 

\begin{equation}
\label{eq:Time-Walk-Fit}
    t_{\text{leading}} = A\cdot\text{ToT}^2 + B\cdot\text{ToT} + C.
\end{equation}
An example of a fitted time-walk curve for a single pixel is shown in Figure \ref{fig:Pixel-Timewalk-Curve}. The time-walk is dependent on the discriminator threshold, therefore the procedure was performed per pixel. Figure \ref{fig:Pixel-Timing-Distribution} shows the timing distribution of a single pixel before and after the time-walk correction was applied. 
The timing performance of the hybrid was evaluated using two methods. The first involves measuring the time-walk corrected timing resolution individually for each pixel and calculating arithmetic mean of the measurements, this defines the ``pixel average'' timing resolution. Figure \ref{fig:Timing-Resolution-Map-200V} shows a 2D distribution of the timing resolution across the pixel matrix and Figure \ref{fig:Matrix-Pixel-Time-Resolution} displays the associated 1D distribution. This method provides information about the variation in pixel performance across the matrix and will be primarily used in this study. To mitigate the impact from pixels with low statistics, typically from those on the periphery of the beam, a cut on the minimum number of hits per pixel is employed, arbitrarily set at \num{500} hits. This approach minimises the impact of low statistics pixels but has limitations. For example, the poor performing pixel in Figure \ref{fig:Pixel-Average-Timing-Resolution-200V}, with a timing resolution of \qty{\sim260}{\ps}, corresponds to a pixel with \num{\sim1000} entries. Whilst a proper time-walk correction is derived and applied the underlying statistical limitation results in a poor extracted timing resolution. However, this produces minimal change in the pixel average timing resolution\footnote{Excluding the entry resulted in negligible change in the pixel average timing resolution and produced an \qty{18}{\percent} reduction in the associated standard deviation.}. Therefore, it was decided to include the entry as similar pixel behaviour is not uncommon in real detector operations. For example, due to pixel inefficiencies or the physical pixel location with respect to the beam spot. 
\paragraph{}
The second method to estimate the hybrid timing performance aggregates the time-walk corrected timestamps from all pixels into a single histogram, an example is shown in Figure \ref{fig:Matrix-Time-Resolution}. The standard deviation of this distribution, extracted from a Gaussian fit, defines the ``matrix'' timing resolution. Whilst, the pixel average measurement will be primarily used, the matrix timing resolution serves as a useful validation. 
\paragraph{}
In both cases, these values include contributions from the \ac{MCP} and the time-stamping chain, as described in Section \ref{subsec:Telescope-Integration-Scheme}. The measured contribution from the time-stamping chain was evaluated to be \qty{55}{\ps}, using test charge injections to the front-end circuit. The difference between the measured and expected contribution of \qty{41}{\ps} is likely due to unaccounted contributions from the internal TDCpix injection circuit. Hereinafter, the hybrid timing resolution refers to the value where the \ac{MCP} and time-stamping chain contributions have been subtracted in quadrature, unless stated otherwise.

\subsection{Timing Resolution vs Sensor Bias}
\label{subsec:Time-Resolution-vs-Bias}
This section presents the hybrid timing resolution as a function of the applied bias voltage to the sensor. The hybrid was operated with a fixed discriminator threshold for all pixels of \qty{1.6}{\femto\coulomb}. As shown in Figure \ref{fig:Hybrid-Time-Resolution}, the timing performance improves with increasing bias voltage, saturating at roughly \qty{110}{\ps} for an applied bias voltage of \qty{250}{\V}. The timing resolution is compared to data collected by NA62 for a single \ac{GTK} station \cite{NA62GTKPaper}, which demonstrates similar performance at high bias voltages. It is highlighted that the \ac{GTK} stations use sensors produced by a different manufacturer and are operated at temperatures below \qty{5}{\degreeCelsius}, compared to the \acp{DUT} which were operated at \qty{\sim20}{\degreeCelsius}. The low operating temperature of the \ac{GTK} allows the required timing resolution to be achieved at a lower operating bias voltage and reduces the impact of accumulated radiation damage.
\paragraph{}

\begin{figure}[h]
    \centering
    \includegraphics[width=0.77\linewidth]{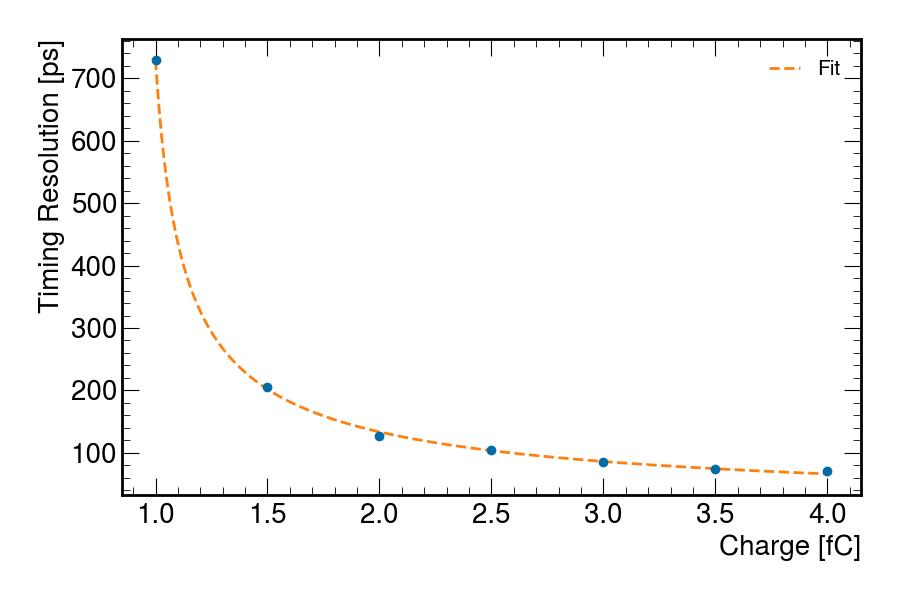}
    \caption{RMS timing resolution $\tau(q)$ of the front-end electronics for a single pixel, determined using test charge injections at a threshold of \qty{1.0}{\femto\coulomb}, fit with a reciprocal curve defined in equation \ref{eq:Reciprocal-Fit}.}
    \label{fig:Chip-Timing-Resolution}
\end{figure}
\begin{figure}
    \centering
    \includegraphics[width=0.77\linewidth]{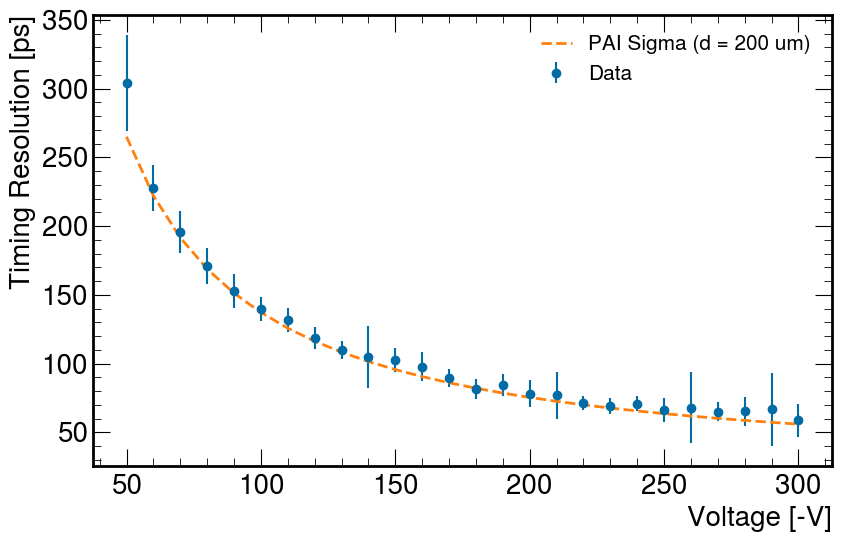}
    \caption{The sensor timing resolution (\ac{ASIC} and \ac{MCP} contributions subtracted), calculated from the pixel average value, as a function of the applied bias voltage. The data is compared with the PAI sigma model (for T = \qty{300}{\K}) presented in \cite{TimeResolutionOfSiliconPixelSensors}.}
    \label{fig:Sensor-Time-Resolution}
\end{figure}

The timing resolution of the hybrid can be separated into two contributions, the sensor and the \ac{ASIC}. The timing resolution of TDCpix was estimated using a series of fixed charge injections to the analogue front-end, with the procedure performed for each pixel in the matrix with a discriminator threshold of \qty{1.0}{\femto\coulomb}\footnote{The fixed charge injections were also performed for a threshold of \qty{1.6}{\femto\coulomb} and minimal effect on the result was seen.}. Furthermore, the sensor was biased to \qty{5}{\V} during the test injections and the hybrid was kept at a temperature of roughly \qty{20}{\degreeCelsius}. Figure \ref{fig:Chip-Timing-Resolution} displays the RMS timing resolution for a single pixel, where the dependence on the registered charge is apparent, fit with a function of the form
\begin{equation}
\label{eq:Reciprocal-Fit}
y = \frac{b^c}{(x -a)^c}.
\end{equation}
Here, $a$, $b$ and $c$ are all free parameters to be determined. The RMS pixel jitter was measured using equipment identical to that used during the testbeam, therefore common contributions from the injection circuit, the sensor and the \ac{TDC} quantisation are included. During this test, the RMS pixel jitter was found to be \qty{108}{\ps} for an input charge of \qty{2.4}{\femto\coulomb}. A similar procedure was previously performed with demonstrator TDCpix modules, without a sensor attached, which found the RMS pixel jitter to be \qty{65}{\ps} for the same input charge and a threshold of \qty{0.9}{\femto\coulomb} \cite{TDCpixPaper}. These results are not directly comparable due to the difference in testing setup and conditions, particularly the contribution from capacitive effects introduced by the sensor. The \ac{ASIC} contribution $\tau_{\text{TDCpix}}$ was evaluated on a run-by-run basis, using the weighted average 

\begin{equation}
\label{eq:Weighted-Time-Resolution}
    \tau_\text{TDCpix} = \frac{\sum_{k}(n_k \cdot \tau(q_k))}{\sum_{k} n_k},    
\end{equation}
where $n_k$ is the number of entries with registered charge $q_k$ and $\tau(q_k)$ is the associated timing resolution, estimated using the aforementioned test injections. Typically, $\tau_{\text{TDCpix}}$ took a value of \qty{\sim 100}{\ps}. Then, the \ac{ASIC} contribution was subtracted in quadrature from the hybrid timing resolution to obtain the sensor timing resolution. The dependence of the latter on the applied bias voltage is shown in Figure \ref{fig:Sensor-Time-Resolution}, and is displayed alongside the ``PAI sigma'' model described in \cite{TimeResolutionOfSiliconPixelSensors}.
This model describes the timing resolution of a parallel plate silicon sensor operated at a large over-depletion voltage, using the \ac{PAI} model for energy loss \cite{PAIModelAllisonandCobb}. The model calculates the timing resolution using a Gaussian fit to the centroid time distribution, similar to the procedure used in this study. Whilst the model assumes a simplified detector geometry, the data displays good agreement with the prediction. 

\section{Integration with TimePix4 Tracks}
\label{sec:TimePix4-Track-Integration}
The tracks reconstructed with the TimePix4 telescope provided an opportunity to investigate the intra-pixel performance of the TDCpix hybrid. This section outlines the procedure for matching TimePix4 tracks to TDCpix hits.

\subsection{Time Alignment}
\label{subsec:Time-Alignment}
\begin{figure}
    \centering
    \includegraphics[width=0.6\linewidth]{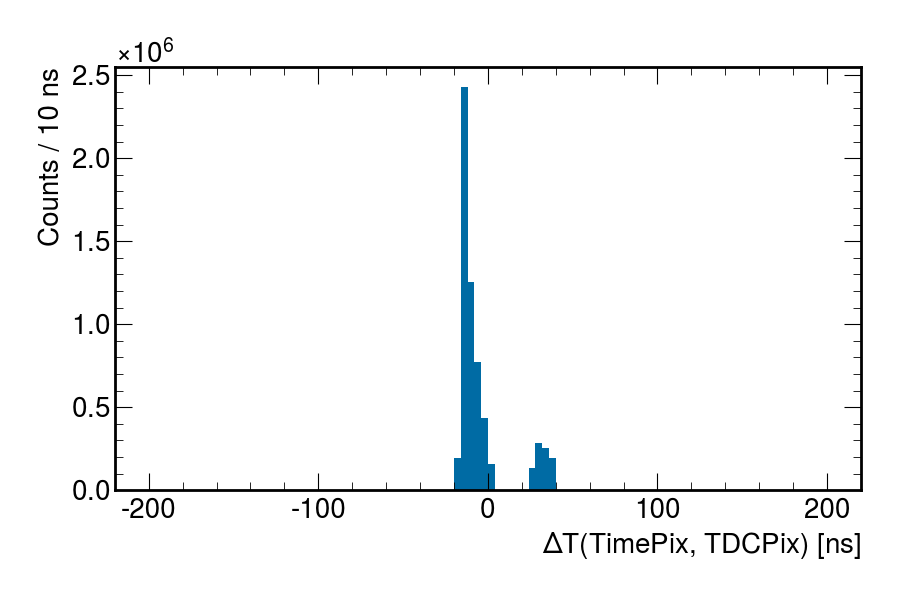}
    \caption{Example of the time difference between the corrected digital TDCpix timestamp and matched TimePix4 track timestamp after the temporal association algorithm. The second peak at \qty{\sim35}{\ns} is from a later spill in the run.}
    \label{fig:Temporal-Track-Matching}
\end{figure}

The first stage of track matching was temporally synchronising the TimePix4 tracks and TDCpix data-stamps. As both TimePix4 and TDCpix recorded timing information from the telescope \ac{MCP} and were aligned to within $\mathcal{O}$(1~s), through the common start-of-run reset signal, synchronisation was achieved by applying the following transformation

\begin{equation}
\label{eq:Time-Alignment-Transformation}
    t_{\text{corr}} = c_{\text{clock}} \cdot t_{\text{TDCpix}} + t_{\text{fine}}.
\end{equation}
Here, $t_{\text{corr}}$ is the corrected TDCpix timestamp, $t_{\text{TDCpix}}$ is the TDCpix digital timestamp (the signal received from the \ac{MCP}), $t_\text{fine}$ is a fine time offset, and $c_{\text{clock}}$ is a multiplicative correction factor. The multiplicative correction factor is necessary to account for small variations in the \ac{DLL} clock frequency which impact the \ac{TDC} binning. The parameter values $c_{\text{clock}}$ and $t_\text{fine}$ were calculated run-by-run by minimising the squared difference $S$ between the TDCpix and TimePix4 timestamps, the minimisation was performed using the SciPy \cite{2020SciPy-NMeth} implementation of the differential evolution algorithm \cite{DifferentialEvolution}. $S$ is explicitly defined as

\begin{equation}
    S(c_{\text{clock}}, t_{\text{fine}}) = \sum_{i = 0}^{N} (t_{\text{corr},~i} - t_{\text{TimePix},~i})^2 = \sum_{i = 0}^{N} (c_{\text{clock}} \cdot t_{\text{TDCpix},~i} + t_{\text{fine}} - t_{\text{TimePix},~i})^2.
\end{equation}
Here, $t_{\text{TimePix}}$ is the corresponding TimePix timestamp and $N$ is the number of timestamps to pair. The corrected TDCpix and TimePix timestamps were paired using a two-pointer technique. This involves forward iterating through the corrected TDCpix and TimePix timestamps; if the time difference $\Delta_t= t_{\text{TDCpix}} - t_{\text{TimePix}}$ between the two selected timestamps are within a pre-specified window $w$ then they are considered matched and the pointers are advanced to the next respective element. If $\Delta_t < -w$ then the TDCpix pointer is advanced, otherwise the TimePix4 pointer is advanced. Due to inefficiencies in both the track reconstruction algorithm and the TDCpix hybrid performance, a small percentage of timestamps remained unpaired; however, this did not significantly impact the analysis. The time difference between matched TDCpix and TimePix4 timestamps is shown in Figure \ref{fig:Temporal-Track-Matching}. The second peak in the distribution corresponds to a different spill in the run, as $c_{\text{clock}}$ varies between spills. The stability of the clock correction was determined by calculating $c_{\text{clock}}$ for each spill and comparing to the value found for the whole run. For the data in Figure \ref{fig:Temporal-Track-Matching}, after converting $c_{\text{clock}}$ back to frequency, the \qty{320}{\MHz} \ac{DLL} clock frequency was determined to be stable within \qty{5}{\Hz}.

\subsection{Spatial Alignment}
\label{subsec:Spacial-Alignment}
\begin{figure}[h]
    \centering
    \begin{subfigure}[t]{0.49\textwidth}
        \centering
        \includegraphics[width=\textwidth]{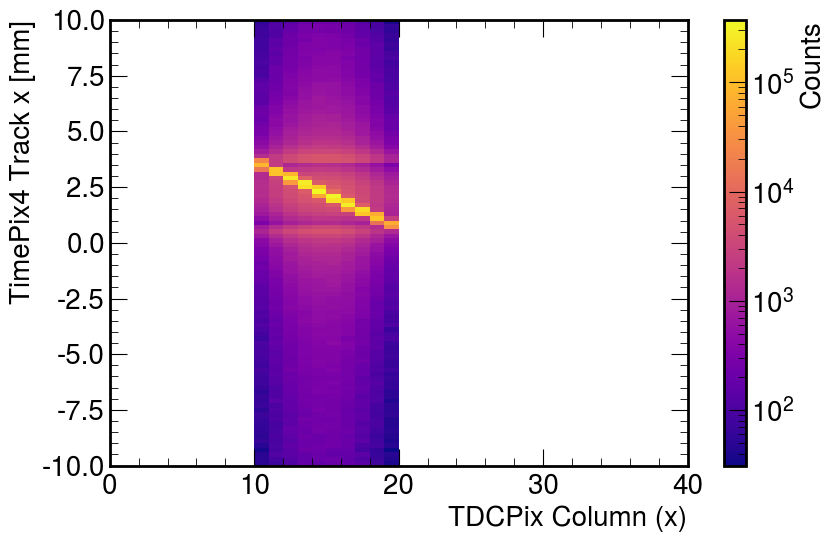}
        \caption{}
        \label{fig:Spatial-Correlation-X}
    \end{subfigure}
    \hfill
    \begin{subfigure}[t]{0.49\textwidth}
        \centering
        \includegraphics[width=\textwidth]{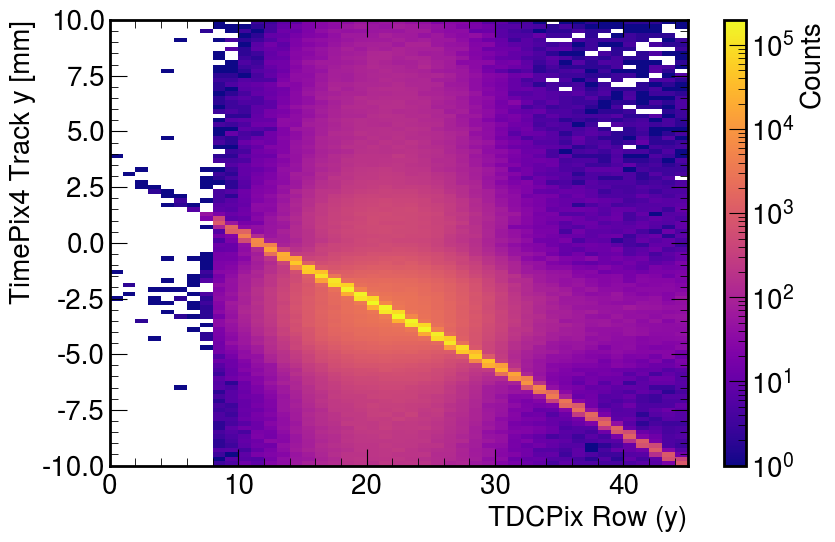}
        \caption{}
        \label{fig:Spatial-Correlation-Y}
    \end{subfigure}
    \caption{Example of the spatial correlation between the TimePix track and the temporally associated TDCpix hit, where only one \acl{QC} was enabled. (a) Between the TimePix track x-coordinate and TDCpix hit column, and (b) between the TimePix track y-coordinate and TDCpix hit row.}
\label{fig:Spatial-Correlation-Plots}
\end{figure}

Once the timestamps were temporally associated, the next stage was to associate the TimePix4 tracks to TDCpix hits. The TimePix tracks were originally specified using four parameters: the track position at $z =0$ ($x_0$, $y_0$) and the track slopes ($\frac{dx}{dz}$, $\frac{dy}{dz}$). Therefore, the tracks were projected onto the TDCpix $z$ position at $z$ = \qty{900}{\mm} to give a pair of Cartesian coordinates ($X_\text{{TimePix}}$, $Y_\text{{TimePix}}$). These coordinates could then be transformed into TDCpix pixel coordinates (row, column) using a linear transformation, as seen from Figure \ref{fig:Spatial-Correlation-Plots}, of the form

\begin{equation}
\label{eq:TimePix4-TDCpix-Correlation}
\begin{aligned}
    \text{Column}_\text{TDCpix} = A \cdot X_\text{TimePix} + B, 
    \\
    \text{Row}_\text{TDCpix} = C \cdot Y_{\text{TimePix}} + D.
\end{aligned}
\end{equation}
Here, $A$ and $C$ are fixed to the inverse of the TDCpix pixel pitch (1 / \qty{300}{\um}), and $B$ and $D$ give the spatial offset between the two coordinate systems. The final step, to extract the values of $B$ and $D$, was to constrain the maximum difference between the TDCpix coordinate and associated TimePix4 coordinate, now in the TDCpix coordinate system, to within one pixel:

\begin{equation}
\label{eq:TimePix4-TDCpix-Correlation-Constraint}
\begin{aligned}
    \left| \text{Column}_\text{TDCpix} - \left(\frac{X_\text{TimePix}}{0.3~\text{mm}} + B\right) \right| \leq 0.5,
    \\
    \left| \text{Row}_\text{TDCpix} - \left(\frac{Y_\text{TimePix}}{0.3~\text{mm}} + D\right) \right| \leq 0.5.
\end{aligned}
\end{equation} 
Due to the nature of the experimental conditions, a small angular misalignment of the \ac{DUT} was expected. The position of the stage that housed the \ac{DUT} box was set to be parallel to the $x$ ($\theta$ rotations) and $y$ ($\phi$ rotations) directions, accurate to within \qty{1.1}{\milli\radian} and \qty{0.6}{\milli\radian} respectively \cite{akiba2025timepix4beamtelescope}. This ensured any misalignment in the \ac{DUT} box was minimised. A misalignment of the hybrid inside the \ac{DUT} box of up to \qty{5}{\degree} would produce a maximum degradation in the spatial resolution of  1 / $\cos{5\degree}=$ \qty{0.4}{\percent}. This contribution is negligible and therefore it is reasonable to conclude that any angular misalignments of the \ac{DUT} had no impact on the outcome of this study.

\section{Intra-Pixel Timing Performance}
\label{sec:Pixel-Timing-Performance}
The procedure described in the previous section enabled the study of the intra-pixel timing resolution. The impact of the weighting field $E_w$ on the timing resolution of a planar pixel detector becomes significant when the pixel pitch $w$ is comparable to the sensor thickness $d$, as discussed in \cite{TimeResolutionOfSiliconPixelSensors}. The TDCpix hybrid provided an excellent opportunity to study the weighting field contribution to the timing performance due to its large pixel pitch and pitch-to-thickness ratio of 1.5. This section details the analysis of the intra-pixel timing performance, comparison with a data-driven model, and comparison with an AllPix$^2$ simulation.

\subsection{Intra-pixel Timing Resolution}
\label{subsec:In-Pixel-Time-Resolution}
\begin{figure}
    \centering
    \begin{subfigure}[t]{0.32\textwidth}
        \centering
        \includegraphics[width=\textwidth]{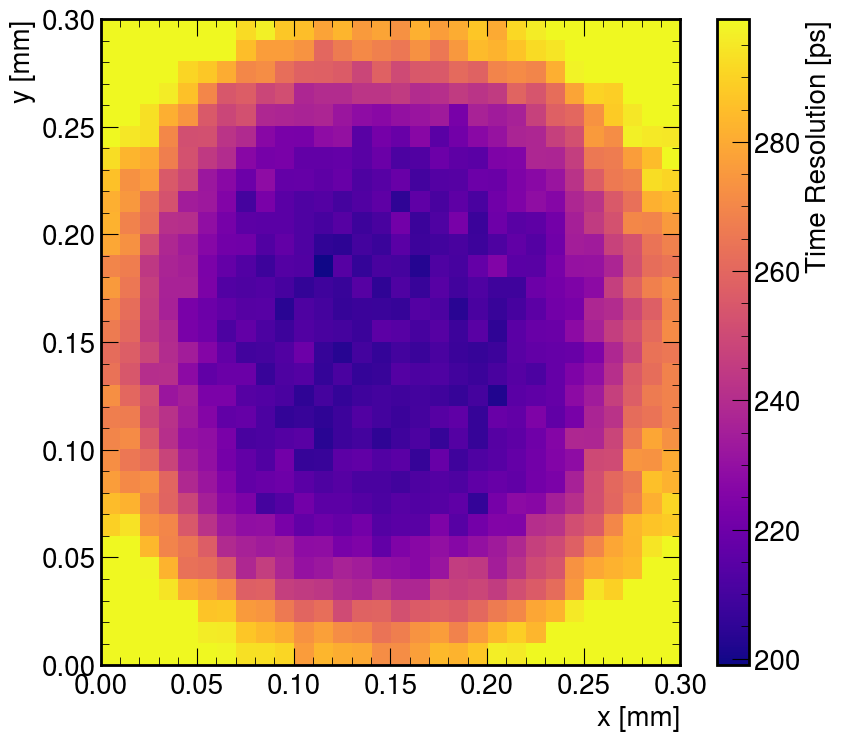}
        \caption{50 V}
        \label{fig:Timing-Resolution-50V}
    \end{subfigure}
    \hfill
    \begin{subfigure}[t]{0.32\textwidth}
        \centering
        \includegraphics[width=\textwidth]{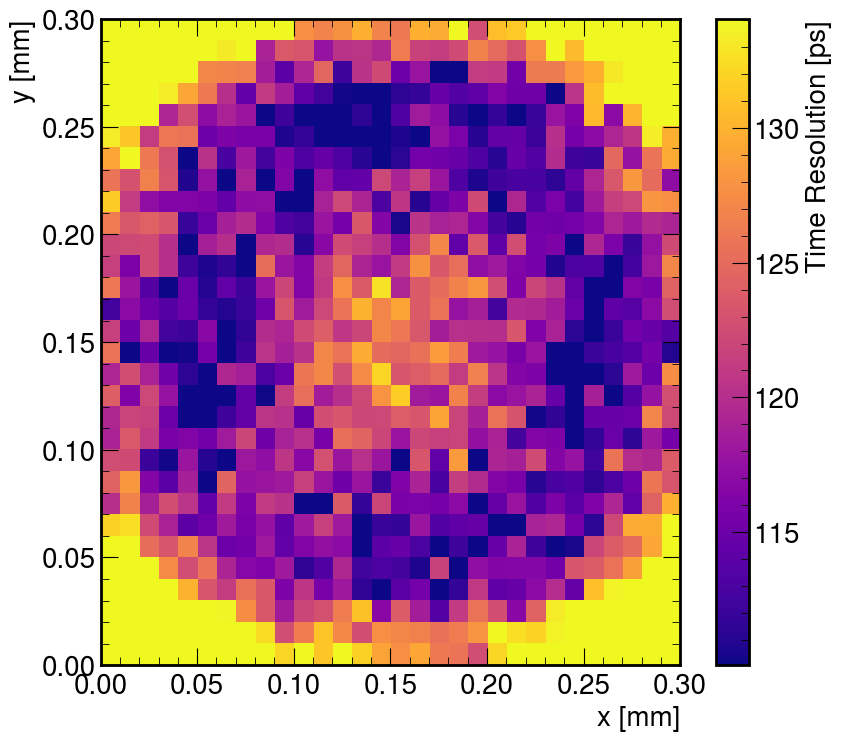}
        \caption{100 V}
        \label{fig:Timing-Resolution-100V}
    \end{subfigure}
    \hfill
    \begin{subfigure}[t]{0.32\textwidth}
        \centering
        \includegraphics[width=\textwidth]{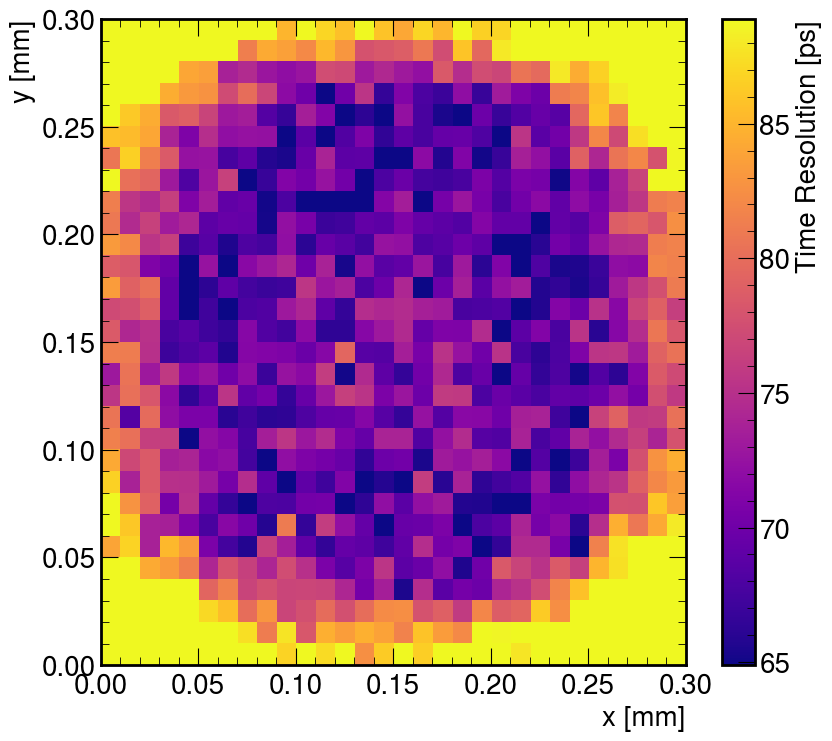}
        \caption{200 V}
        \label{fig:Timing-Resolution-200V}
    \end{subfigure}
    \caption{Evolution of the intra-pixel sensor timing resolution (\ac{ASIC} and \ac{MCP} contributions subtracted) with increasing bias voltage. Time-walk corrections have been applied and measurements from all pixels have been aggregated. The colour scale is not fixed between maps to highlight the individual key features.}
\label{fig:Intra-Pixel-Timing-Resolution-vs-Voltage}
\end{figure}

\begin{figure}
    \centering
    \begin{subfigure}[t]{0.32\textwidth}
        \centering
        \includegraphics[width=\textwidth]{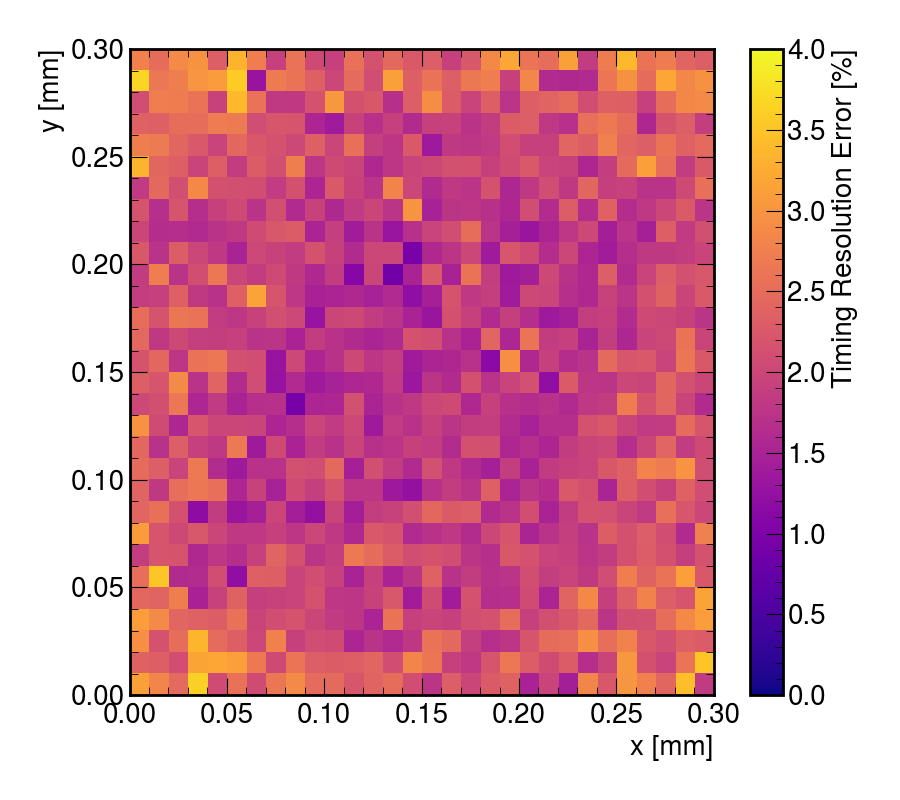}
        \caption{50 V}
        \label{fig:Timing-Resolution-Error-50V}
    \end{subfigure}
    \hfill
    \begin{subfigure}[t]{0.32\textwidth}
        \centering
        \includegraphics[width=\textwidth]{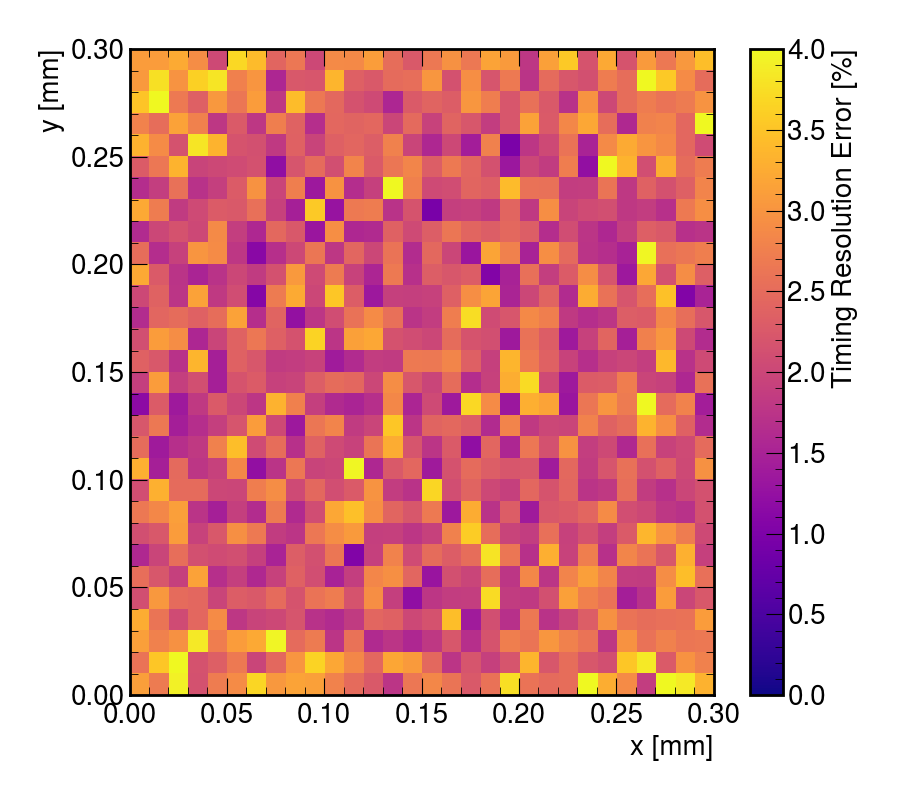}
        \caption{100 V}
        \label{fig:Timing-Resolution-Error-100V}
    \end{subfigure}
    \hfill
    \begin{subfigure}[t]{0.32\textwidth}
        \centering
        \includegraphics[width=\textwidth]{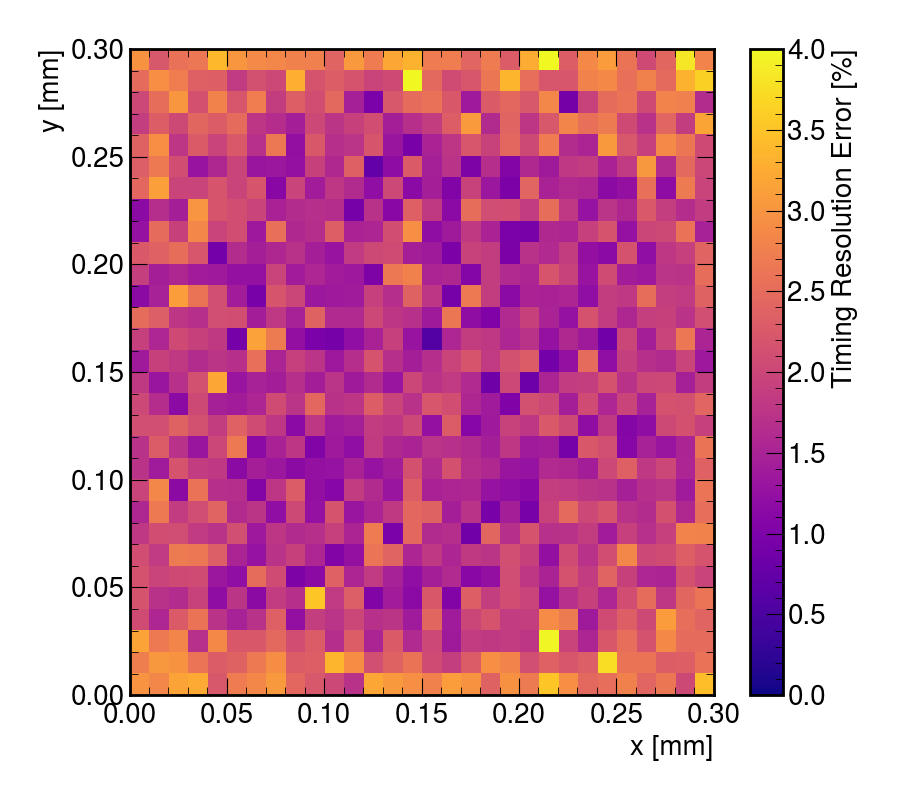}
        \caption{200 V}
        \label{fig:Timing-Resolution-Error-200V}
    \end{subfigure}
    \caption{Spatial variation in the error on the intra-pixel timing resolution (Figure \ref{fig:Intra-Pixel-Timing-Resolution-vs-Voltage}), as reported by the fitting procedure. Each plot is normalized to the average intra-pixel timing resolution for the corresponding bias voltage.}
\label{fig:Intra-Pixel-Timing-Resolution-Error-vs-Voltage}
\end{figure}

The intra-pixel timing resolution was studied as a function of the in-pixel coordinates ($x$, $y$), shown in Figure \ref{fig:Intra-Pixel-Timing-Resolution-vs-Voltage}. The data from all pixels have been aggregated to improve the statistics in each bin, and time-walk corrections have been applied. Here, it is visible that the timing resolution degrades near the periphery of the pixel, potentially due to charge sharing between pixels. In the \qtylist{50;200}{\V} configurations, the best performing region is the centre of the pixel. However, for the \qty{100}{\V} case, the best region is confined to an annulus between the centre and periphery of the pixel.
\paragraph{}
The intra-pixel timing resolution is shown to vary with the applied bias voltage. At bias voltages near the depletion voltage (Figure \ref{fig:Timing-Resolution-50V}) the timing resolution is uniform across the pixel centre; however, as the bias voltage increases the timing resolution becomes increasingly inhomogeneous (Figure \ref{fig:Timing-Resolution-100V}). When the sensor is operated at a large over-depletion voltage the timing performance returns to being uniform across the pixel centre (Figure \ref{fig:Timing-Resolution-200V}). The error on the intra-pixel timing resolution for each voltage is shown in the maps presented in Figure \ref{fig:Intra-Pixel-Timing-Resolution-Error-vs-Voltage}. No correlation is seen with the features apparent in the intra-pixel timing resolution maps, providing evidence that the distributions are a result of a physical mechanism. 


\subsection{Intra-pixel Time of Arrival}
\label{subsec:Intra-Pixel-ToA}

\begin{figure}
    \centering
    \begin{subfigure}[t]{0.32\textwidth}
        \centering
        \includegraphics[width=\textwidth]{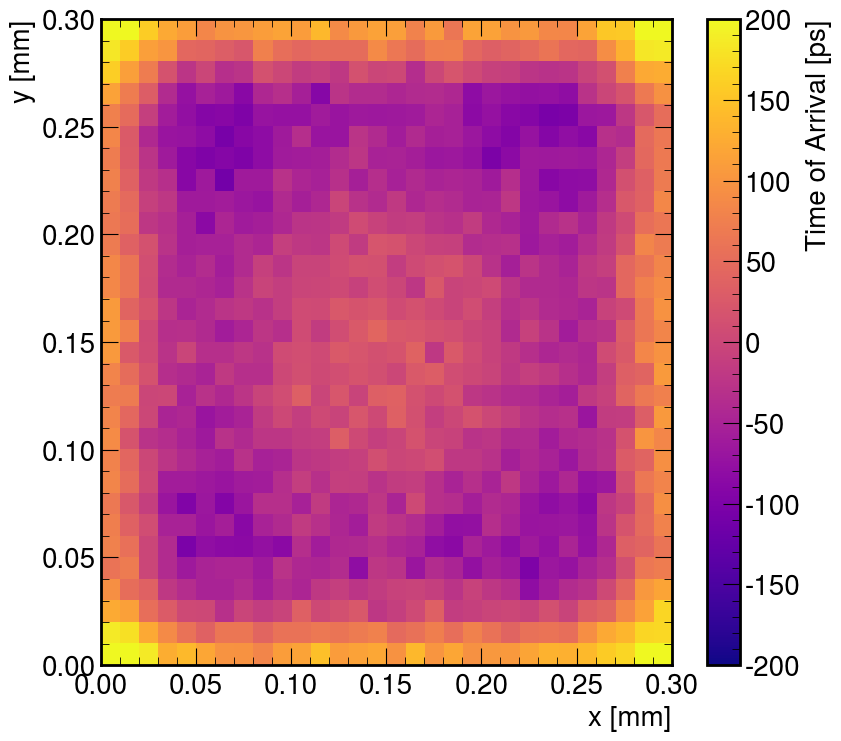}
        \caption{50 V}
        \label{fig:ToA-Map-50V}
    \end{subfigure}
    \hfill
    \begin{subfigure}[t]{0.32\textwidth}
        \centering
        \includegraphics[width=\textwidth]{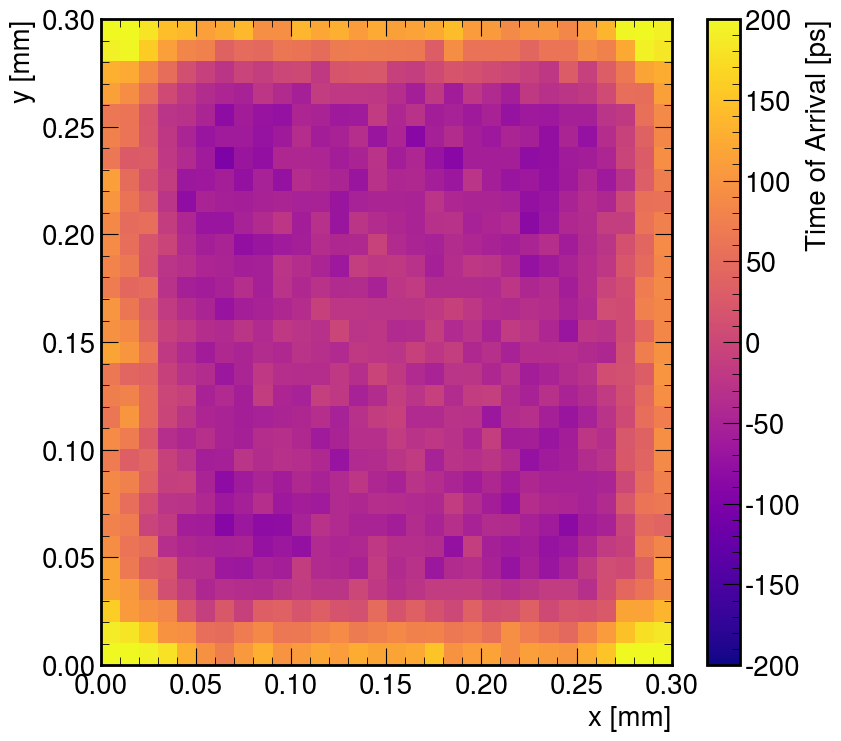}
        \caption{100 V}
        \label{fig:ToA-Map-100V}
    \end{subfigure}
    \hfill
    \begin{subfigure}[t]{0.32\textwidth}
        \centering
        \includegraphics[width=\textwidth]{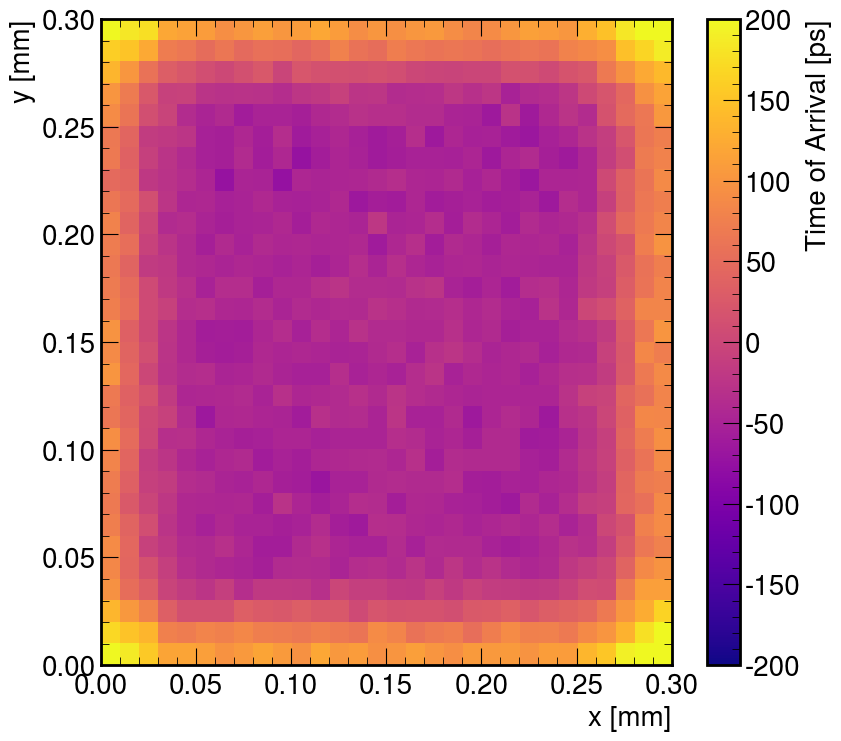}
        \caption{200 V}
        \label{fig:ToA-Map-200V}
    \end{subfigure}
    \caption{Evolution of the intra-pixel mean subtracted \ac{ToA} with increasing bias voltage. Propagation delay corrections have been applied and measurements from all pixels have been aggregated. The colour scale is fixed to [-200, 200]~ps.}
\label{fig:Intra-Pixel-ToA-vs-Voltage}
\end{figure}

The timing resolution is influenced by multiple contributions, such as the time-walk. Therefore it is useful to study the variations in the leading-edge time of the signal, hereinafter referred to as the \ac{ToA}, as it is free from analysis corrections and the associated uncertainties. In practice, the mean subtracted \ac{ToA} is used, where the average \ac{ToA} for the run is subtracted. To aggregate the measurements from all pixels, a per-pixel propagation delay correction is applied due to the different time it takes for a signal to traverse a transmission line to the \ac{EoC} region. The propagation delay is extracted from the mean of the uncorrected timing distribution, evaluated using a Gaussian fit, and subtracted from the timestamps for the given pixel. The resulting correction is constant across the intra-pixel region of a given pixel. Time-walk corrections, which include the pixel propagation delay by construction, are not applied. The \ac{ToA} was similarly studied as a function of the in-pixel coordinates, shown in Figure \ref{fig:Intra-Pixel-ToA-vs-Voltage}. In this case, the areas with the best timing performance are four localised regions near the pixel corners. The intra-pixel \ac{ToA} is also shown to vary with the applied bias voltage, becoming more uniform across the pixel at high bias voltages. 

\subsection{Weighting Field}
\label{subsec:Weighting-Field}
\begin{figure}
    \centering
    \begin{subfigure}[t]{0.49\textwidth}
        \centering
        \includegraphics[width=\textwidth]{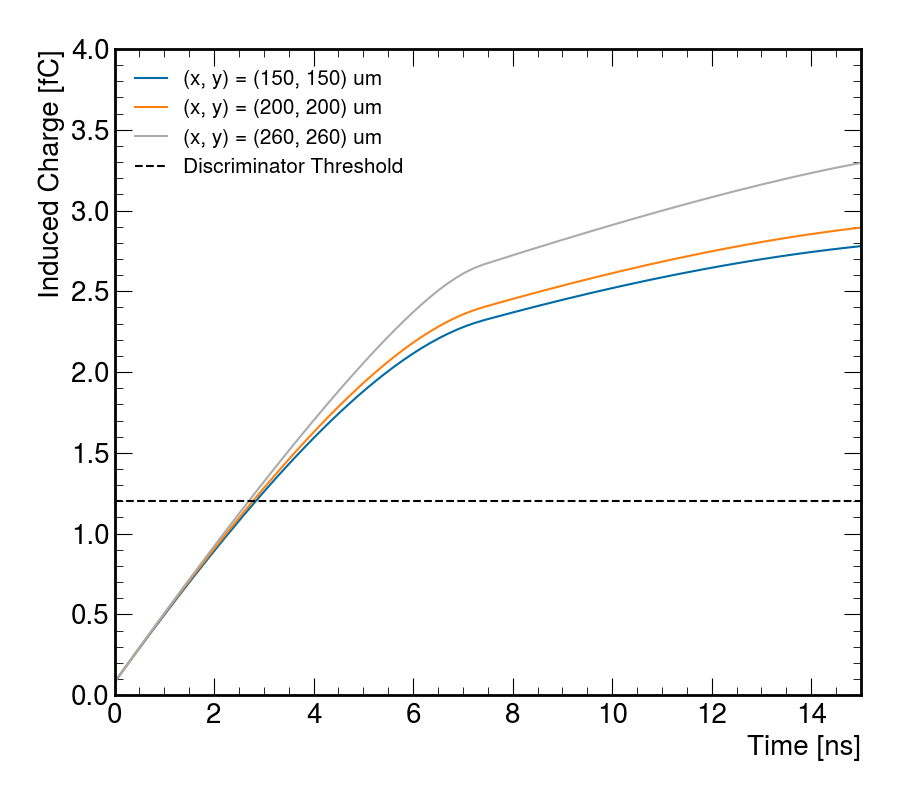}
        \caption{}
        \label{fig:Charge-Pulse-50V}
    \end{subfigure}
    \hfill
    \begin{subfigure}[t]{0.49\textwidth}
        \centering
        \includegraphics[width=\textwidth]{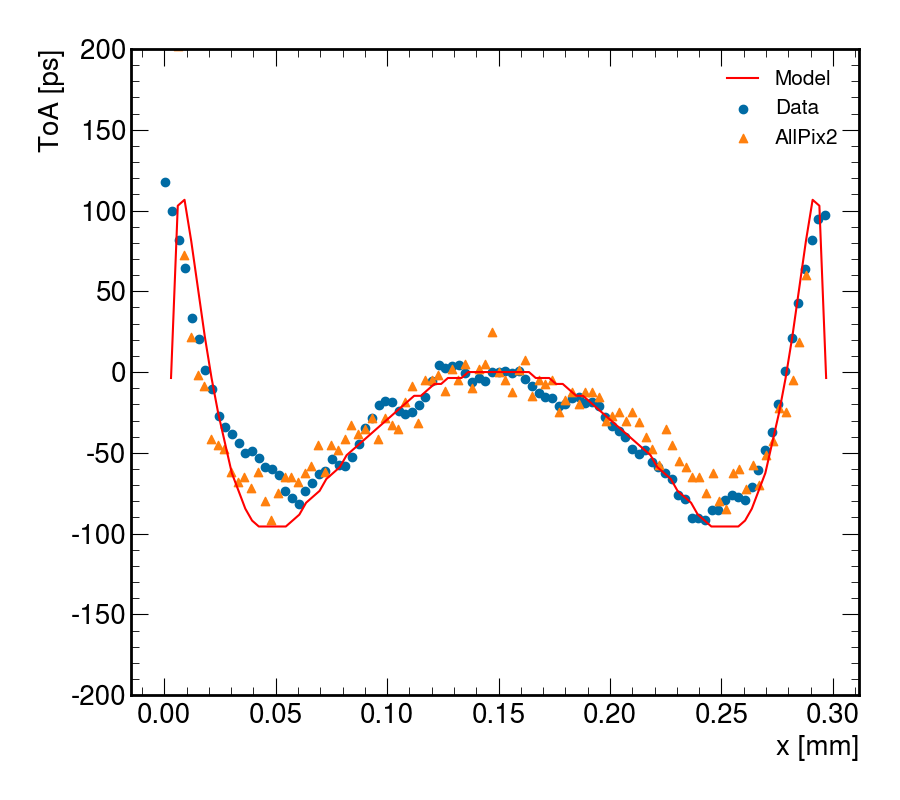}
        \caption{}
        \label{fig:Numerical-ToA-Comparison}
    \end{subfigure}
\caption{(a) Numerically calculated charge pulses for various intra-pixel positions, where $x = y = 0$ is the bottom left pixel corner. The discriminator threshold is indicated by the dashed black line. (b) Comparison between the numerically calculated \ac{ToA} (red), AllPix$^2$ simulation (orange, Section \ref{subsec:AllPix2-Simulation}) and data (blue) mean subtracted \ac{ToA} as a function of the intra-pixel $x$-coordinate, taken at $y$ = \qty{0.15}{\mm}. All values are taken for a bias voltage of \qty{50}{\V}.}
\end{figure}

\begin{figure}
    \centering
    \includegraphics[width=0.6\linewidth]{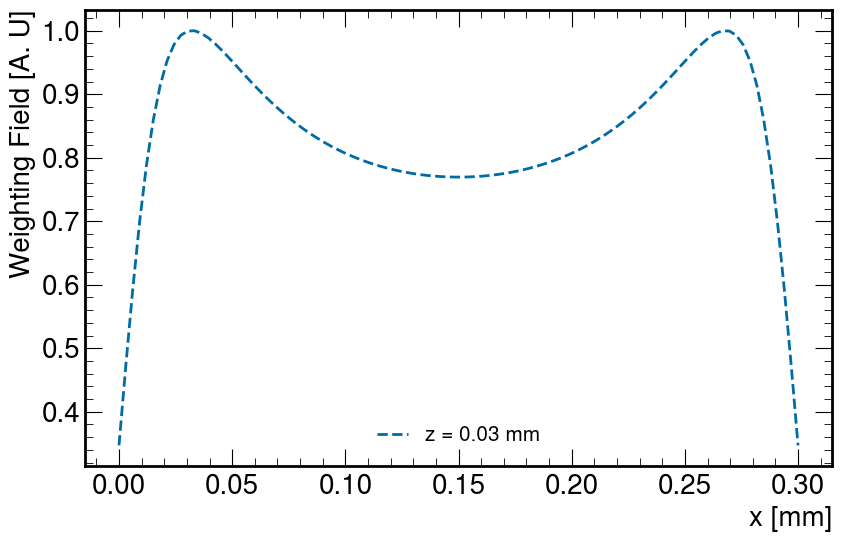}
    \caption{Projection of the weighting field, defined in \cite{TimeResolutionOfSiliconPixelSensors}, at $z$ = \qty{0.03}{\mm}, $y$ = \qty{0.15}{\mm}. The weighting field is maximal closer to the periphery of the pixel and has a local minimum near the pixel centre, correlating with the observed intra-pixel \ac{ToA} projection shown in Figure \ref{fig:Numerical-ToA-Comparison}.}
    \label{fig:Weighting-Field-Projection}
\end{figure}

To understand the observed features, the impact of the weighting field $E_w$ on the \ac{ToA} was investigated using a data-driven numerical model and a set of simplifying assumptions, namely: the electric field $\vec{E}$ is uniform in $x$ and $y$, the charge deposition is fixed and uniform (\qty{73}{\elementarycharge\per\um} \cite{PDG_2006}) along the particle track, the front-end circuit consists of an ideal integrator and discriminator, and the sensor has a parallel plate geometry. The final assumption allows the use of the analytically calculated weighting field found in \cite{TimeResolutionOfSiliconPixelSensors}. The induced current $i_{ind}$ for a single charge, at a position $\vec{r} = (x, y, z)$ in the pixel bulk, was calculated using the Shockley-Ramo theorem \cite{Shockley-Paper, Ramo-Paper} and passed through the ideal front-end circuit to extract the \ac{ToA}. In reality, the analytical weighting field does not account for edge effects, therefore an effective weighting field $E_w^{\text{eff}}(\vec{r})$ was introduced to the calculated induced current 

\begin{equation}
\label{eq:Modified-Shockley-Ramo}
    i_{ind} = q\vec{E_w^{\text{eff}}}(\vec{r}) \cdot \vec{v_{dr}}(E(z), T),
\end{equation}
where $q$ is the charge of the carrier (electron or hole), $\vec{v_{dr}}$ is the drift velocity of the carrier (parametrised as a function of the transverse electric field $E(z)$ and temperature $T$) and $E_w^{\text{eff}}(\vec{r})$ is the effective weighting field. The temperature dependence of the drift velocity was implemented following the parametrisation in \cite{EREMIN2004622}, and the electric field was implemented following the model in \cite{LI1997297}. The effective weighting field is defined as 
\begin{equation}
\label{eq:Effective-Weighting-Field}
    \vec{E_w^{\text{eff}}}(\vec{r}) = \vec{E_w}(\vec{{r}}) \cdot f(x, y),
\end{equation}
where $E_w$ is the analytical weighting field defined in \cite{TimeResolutionOfSiliconPixelSensors} and $f(x, y)$ is the edge effect function, which takes the form
\begin{gather}
\label{eq:Separable-Transport-Equation}
f(x, y) = g(x) \cdot g(y), \\
\text{where}~g(v) = A~\text{erf}\left(\frac{0.5w - v - B}{C}\right) + (1 - A). 
\end{gather}
Here, $w$ is the pixel pitch, and $A$, $B$ and $C$ are empirical parameters. The parameters are calculated using a data-driven approach by minimising the squared difference between the modelled \ac{ToA} and the data. Typical values of the parameters are: $A \in (0, 1)$, $B = \mathcal{O}$(\qty{10}{\um}) and $C = \mathcal{O}$(\qty{10}{\um}). Examples of charge pulses, for various intra-pixel positions, are shown in Figure \ref{fig:Charge-Pulse-50V}, displaying the differences in threshold crossing time. A comparison between the modelled \ac{ToA} and the experimental data, obtained at a bias voltage of \qty{50}{\V}, is shown in Figure \ref{fig:Numerical-ToA-Comparison}. Upon comparison with the expected weighting field for a pixel electrode, a projection is shown in Figure \ref{fig:Weighting-Field-Projection}, it is reasonable to infer that the shape of the \ac{ToA} distribution is correlated with the weighting field and the resultant induced current. 
\begin{figure}
    \centering
    \begin{subfigure}[t]{0.49\textwidth}
        \centering
        \includegraphics[width=\textwidth]{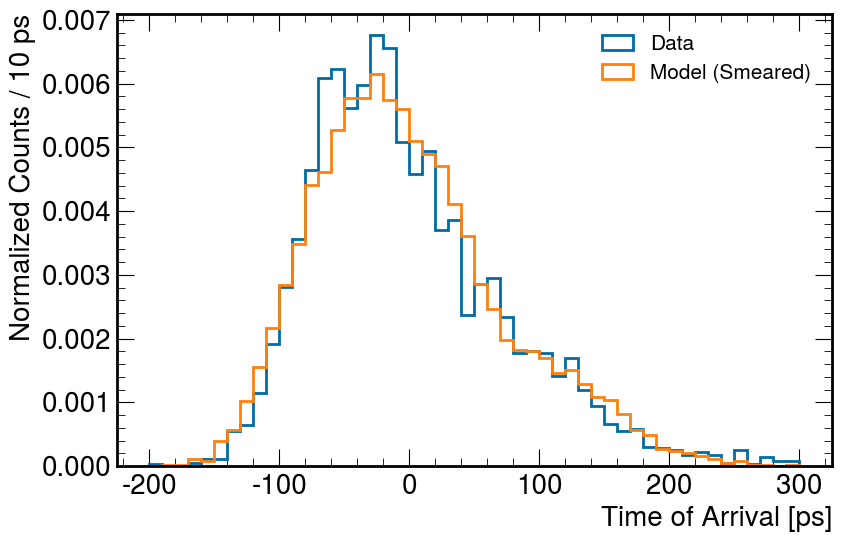}
        \caption{}
        \label{fig:ToA-Comparison-Histogram-50V}
    \end{subfigure}
    \hfill
    \begin{subfigure}[t]{0.49\textwidth}
        \centering
        \includegraphics[width=\textwidth]{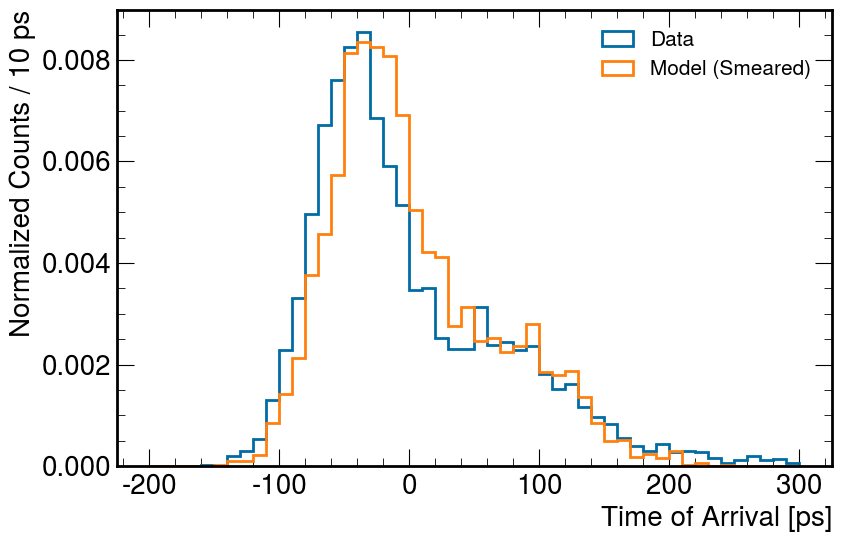}
        \caption{}
        \label{fig:ToA-Comparison-Histogram-100V}
    \end{subfigure}
\caption{Distributions of the intra-pixel \ac{ToA}, for data (blue) and the numerical model (orange), at (a) \qty{50}{\V} and (b) \qty{100}{\V} bias voltage. A Gaussian smearing of \qty{28}{\ps} has been applied to the model values to simulate the effect of the TDCpix \ac{TDC} binning. A systematic offset of \qty{+50}{\ps} was added to the \qty{50}{\V} model distribution, to improve agreement with the data. No offset was added to the \qty{100}{\V} distribution. The distributions are normalised to unity and the range restricted to [-200, 300]~ps.}
\label{fig:ToA-Comparison-Histograms}
\end{figure}
A comparison between the modelled and experimental \ac{ToA} distributions, for the whole pixel, at different bias voltages is shown in Figure \ref{fig:ToA-Comparison-Histograms}. Here, a Gaussian smearing of \qty{97.7}{\ps} $/ \sqrt{12} = $ \qty{28}{\ps} has been applied to the model to simulate the impact of the \ac{TDC} binning. The model qualitatively displays good agreement with the intra-pixel ToA distributions, across a range of bias voltages. The spatial variation in the \ac{ToA}, arising from the weighting field, contributes to the timing resolution of the sensor. The standard deviations of the \ac{ToA}, for the data and the model, is listed alongside the sensor timing resolution in Table \ref{tab:ToA-Timing-Resolution-Comparison}. The error on $\sigma(\text{ToA})$ is a systematic uncertainty, estimated from varying the intra-pixel binning.  It is apparent that as the bias voltage is increased, $\sigma$(\ac{ToA}) approaches the timing resolution. Therefore, it is reasonable to conclude that, at high bias voltages, the timing resolution of the sensor is dominated by the weighting field contribution. Importantly, the point at which the weighting field effect becomes dominant (contributes to \qty{\sim 50}{\percent} of the overall timing resolution) is for $V_b$ \qty{>= 200}{\V}, an operating range relevant for silicon devices.

\begin{table}[]
\centering
\caption{Comparison of the standard deviation of the numerically calculated and data \ac{ToA} with the sensor timing resolution. The model values are obtained for a charge deposition of \qty{2}{\femto\coulomb} and a threshold set at \qty{1.16}{\femto\coulomb}. The sensor timing resolution is calculated using the pixel average method.}
\begin{tabular}{|c|c|c|c|}
\hline
Voltage [-V] & $\sigma$(ToA) Model [ps] & $\sigma$(ToA) Data [ps] & Sensor $\sigma_t$ [ps] \\ \hline
50      & $74$            & $73 \pm 6$         & $256 \pm 15$                  \\ \hline
100     & $64$           & $69 \pm 5$         & $129 \pm 3$                   \\ \hline
200     & $64$            & $65 \pm 5$         & $83 \pm 2$                    \\ \hline
\end{tabular}
\label{tab:ToA-Timing-Resolution-Comparison}
\end{table}

\subsection{AllPix$^2$ Simulation}
\label{subsec:AllPix2-Simulation}
\begin{figure}
    \centering
    \begin{subfigure}[t]{0.32\textwidth}
        \centering
        \includegraphics[width=\textwidth]{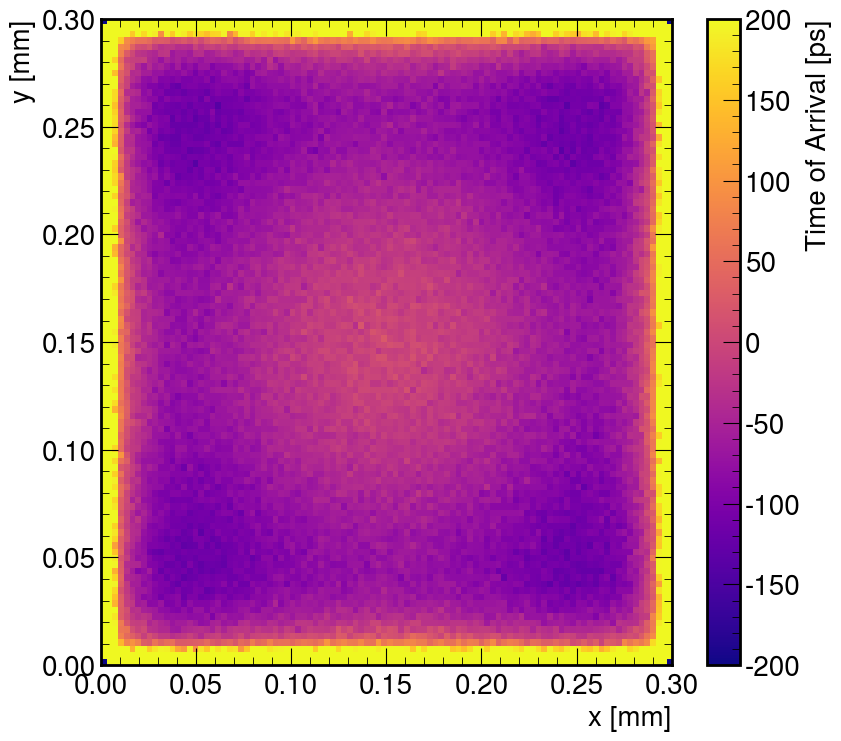}
        \caption{50 V}
        \label{fig:AllPix-ToA-Map-50V}
    \end{subfigure}
    \hfill
    \begin{subfigure}[t]{0.32\textwidth}
        \centering
        \includegraphics[width=\textwidth]{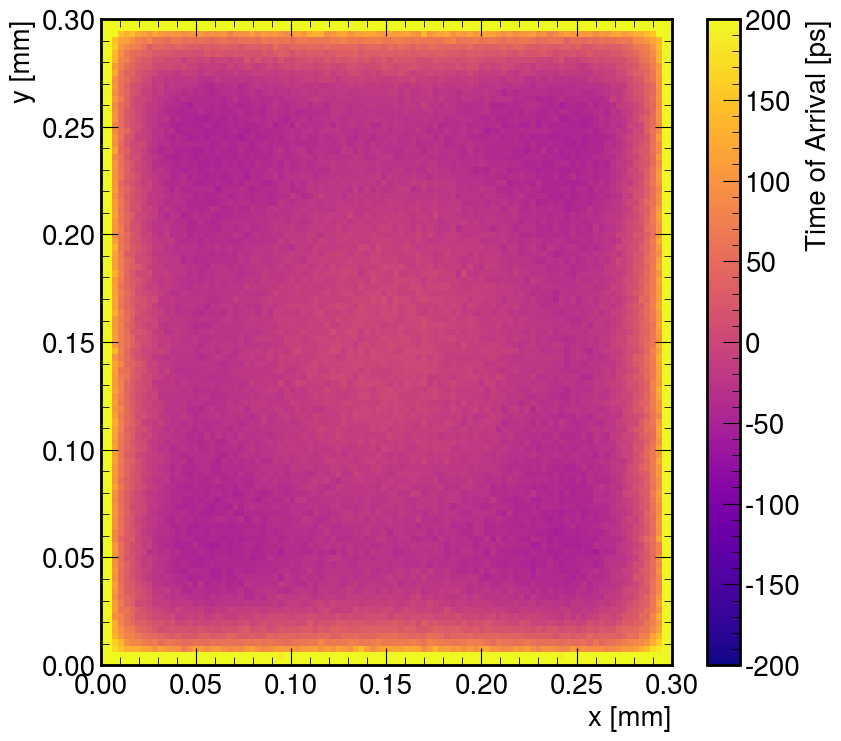}
        \caption{100 V}
        \label{fig:AllPix-ToA-Map-100V}
    \end{subfigure}
    \hfill
    \begin{subfigure}[t]{0.32\textwidth}
        \centering
        \includegraphics[width=\textwidth]{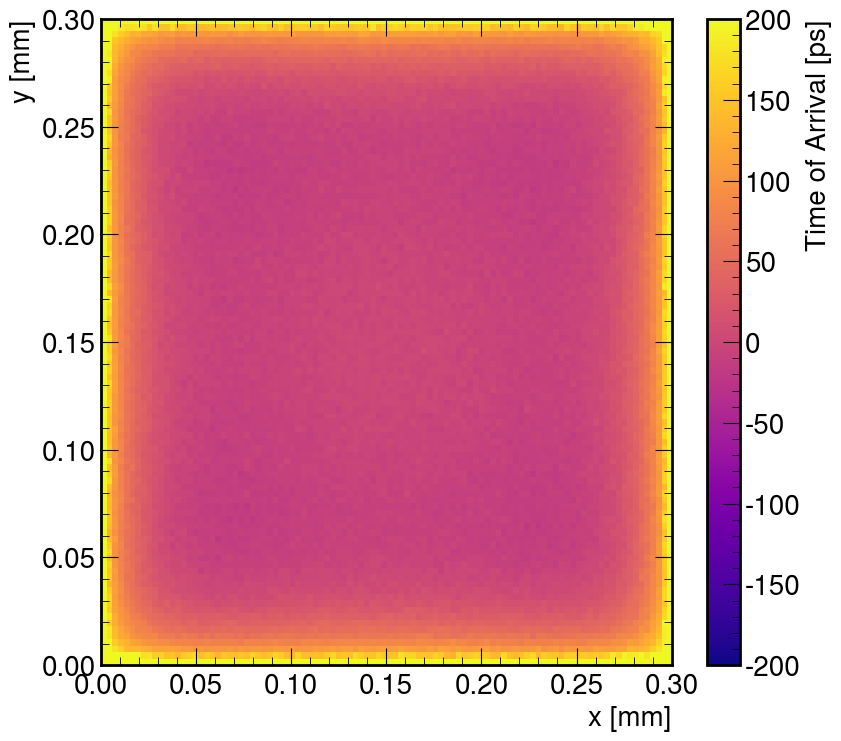}
        \caption{200 V}
        \label{fig:AllPix-ToA-Map-200V}
    \end{subfigure}
    \caption{\ac{ToA} maps produced from an AllPix$^2$ simulation for different bias voltages. Each map is produced from $1\times10^{5}$ \acp{MIP} with a constant charge deposition. The colour scale is fixed to [-200, 200]~ps and offset such that the \ac{ToA} at $x$ = $y$ = \qty{0.15}{\mm} is zero.}
\label{fig:AllPix-ToA-vs-Voltage}
\end{figure}

An AllPix$^2$ \cite{AllPix2} simulation was performed to validate the findings from the data and data-driven model. The \ac{DUT} was implemented as a simple \qty{200}{\um} thick ``pixelated monolithic'' sensor using in-built AllPix$^2$ models. The charge deposition, propagation, pulse transfer and digitisation were all performed using in-built modules. The simulation used a pre-calculated form of the weighting field from \cite{TimeResolutionOfSiliconPixelSensors} and assumes a constant electric field across the sensor volume. One hundred thousand simulated \acp{MIP} were uniformly distributed across the sensor surface, each with a constant charge deposition of \qty{73}{\elementarycharge\per\um} \cite{PDG_2006} in the bulk. The simulation was performed for \qtylist{50;100;200}{\V} bias voltages and the corresponding \ac{ToA} maps are shown in Figure \ref{fig:AllPix-ToA-vs-Voltage}. Notably, four regions of relatively faster performance are seen near the pixel corners, and the \ac{ToA} becomes increasingly homogeneous as the bias voltage increases. The AllPix$^2$ model serves as a useful characterisation tool for the bulk of the sensitive volume, providing a qualitatively accurate description of the pixel timing performance. However, the model breaks down near the pixel edges, as seen in Figure \ref{fig:AllPix-ToA-vs-Voltage}. This is where the data-driven model can be implemented to quantify the edge effects. 

\section{Conclusion}
The timing resolution of the TDCpix hybrid was established and the sensor contribution successfully decoupled and compared with theoretical expectations. Utilising the good spatial resolution of the TimePix4 telescope, the intra-pixel timing resolution has been studied for the first time in an experimental environment. Non-uniformities in the intra-pixel timing resolution have been investigated using the \ac{ToA}, in the context of the weighting field, and compared with a data-driven model and an AllPix$^2$ simulation. Results show significant spatial variations in the intra-pixel timing resolution due to the weighting field. The AllPix$^2$ framework and data-driven models have been validated for planar technologies, with large pixel pitches. Further studies and detailed simulations would be necessary to fully characterise the impact of the weighting field on the intra-pixel timing resolution, particularly for non-planar technologies. 
\paragraph{}
These results provide important context for future 4D silicon-based tracking \ac{RD} efforts. Specifically, those requiring low mass, high granularity systems with a single hit timing resolution \qty{<=100}{\ps}, as the weighting field impact is maximal for a pitch-to-thickness ratio close to unity. These requirements have been identified as key targets for future accelerators and flavour experiments \cite{ECFARoadmap}. For example, the vertex detector for the proposed Muon Collider would need to achieve a single hit timing resolution of \qty{30}{\ps}, with a granularity of \qtyproduct[product-units=single]{25 x 25}{\um\squared} \cite{MuonColliderArticle}. One technology under consideration are \ac{MAPS} with a thickness of $\approx$ \qty{50}{\um}, in this case the impact of the weighting field would have to be considered.
\paragraph{}
In this work, it has been demonstrated that high granularity telescopes can be used to quantify the intra-pixel timing non-uniformities in silicon pixel detectors in conditions that closely mimic the target application environment. The techniques presented can be readily extended to non-planar devices such as: \aclp{LGAD}, \acl{MAPS} and 3D sensors. Similarly, these techniques can be used in tandem with simulation and laboratory studies to fully quantify the intra-pixel timing performance.

\acknowledgments
The authors are thankful to: the NA62 collaboration, in particular the \ac{GTK} working group; the TDCpix development team; the TimePix4 telescope team; and  Che Sum Tse. 
\paragraph{}
C. Sam and A. Shepelev have received support from the Science and Technology Facilities Council (STFC), funding UKRI/ST/Y509036/1 and UKRI/ST/C002848/1, respectively. B. Velghe has received support from the Natural Sciences and Engineering Research Council of Canada (NSERC), funding SAPPJ-2020-0026.


\bibliographystyle{JHEP}
\bibliography{biblio.bib}






\end{document}